\documentclass{iopjournal}
\usepackage{graphicx}
\usepackage{bm}
\usepackage{amssymb}
\usepackage{color}
\usepackage{amsmath}
\usepackage{amstext}
\usepackage{latexsym}
\usepackage{tikz}
\usepackage{ulem}
\usepackage{comment}
\usepackage[maxbibnames = 10, eprint = true, style = phys]{biblatex}
\usepackage[colorlinks=true,citecolor=purple,linkcolor=red,urlcolor=purple]{hyperref}
\usepackage{bbold}
\usepackage{float}
\usepackage{enumerate}
\usepackage{braket}
\usepackage{upgreek}

\definecolor{pink}{rgb}{1,0.078,0.57}

\definecolor{green}{rgb}{0,0.7,0.9}
\usepackage{amsfonts}
\usepackage{mathrsfs}

\newcommand{\beq}{\begin{equation}}
\newcommand{\eeq}{\end{equation}}
\newcommand{\beqa}{\begin{eqnarray}}
\newcommand{\eeqa}{\end{eqnarray}}

\newcommand{\av}[1]{\langle #1 \rangle}

\newcommand{\avs}[1]{\mathbb{E}(#1)} 
\usepackage{tikz}

\newcommand{\be}[1]{\mathbf{#1}}
\newcommand{\mc}[1]{\mathcal{#1}}
\newcommand{\mr}[1]{\mathrm{#1}}
\newcommand{\ts}[1]{\textsc{#1}}

\newcommand{\ha}[1]{\hat{#1}}
\newcommand{\ti}[1]{\tilde{#1}}

\newcommand{\mbb}[1]{\mathbb{#1}}
\newcommand{\bs}[1]{\boldsymbol{#1}}

\newcommand{\dg}{^{\dagger}}
\newcommand{\Tr}{\mathrm{Tr}}
\newcommand{\vrho}{\hat{\varrho}}
\newcommand{\uvrho}{\tilde{\varrho}}

\newcommand{\uL}{\tilde{\mathcal{L}}}

\newcommand{\st}[1]{\MakeUppercase{#1}}

\graphicspath{{./images/}}

\newcommand{\Id}{\hat{\mathbb{1}}}

\makeatletter
\newif\iftag@here

\newcommand*{\taghere}[1][0pt]
{\ifmeasuring@\else
  \global\tag@heretrue
  \tikz[remember picture,overlay]{\coordinate (taghere) at (0pt,#1);}%
\fi}

\def\place@tag{%
    \iftagsleft@
      \kern-\tagshift@
      \iftag@here
        \global\tag@herefalse
        \tikz[remember picture,overlay]%
          {\path (taghere) -| node[anchor=base]{\rlap{\boxz@}} (0pt,0pt);}%
      \else
        \if1\shift@tag\row@\relax
            \rlap{\vbox{%
                \normalbaselines
                \boxz@
                \vbox to\lineht@{}%
                \raise@tag
            }}%
        \else
            \rlap{\boxz@}%
        \fi
        \kern\displaywidth@
      \fi
    \else
      \kern-\tagshift@
      \iftag@here
        \global\tag@herefalse
        \tikz[remember picture,overlay]%
          {\path  (taghere) -|  node[anchor=base]{\llap{\boxz@}} (0pt,0pt);}%
      \else
        \if1\shift@tag\row@\relax
            \llap{\vtop{%
                \raise@tag
                \normalbaselines
                \setbox\@ne\null
                \dp\@ne\lineht@
                \box\@ne
                \boxz@
            }}%
        \else \llap{\boxz@}%
        \fi
      \fi
    \fi
}
\makeatother

\usepackage{fontawesome}
\newcommand{\github}[1]{\href{#1}{\faGithubSquare}}
\newcommand{\esrgithub}{\github{https://github.com/fame64/NonHermitianMagic}}

\begin{document}

\title{Magic Steady State Production:\\
Non-Hermitian, Dissipative, and Stochastic Pathways}

\author{Pablo Martinez-Azcona$^{2, 1, 3, 4,*}$\orcid{0000-0002-9553-2610}, Matthieu Sarkis$^1$\orcid{0009-0002-5494-8406}, Alexandre Tkatchenko$^1$\orcid{0000-0002-1012-4854} and Aur\'elia Chenu$^{1}$\orcid{0000-0002-4461-8289}}

\affil{$^1$Department of Physics and Materials Science, University of Luxembourg, L-1511 Luxembourg}

\affil{$^2$Technical University of Munich, TUM School of Natural Sciences,
Physics Department, 85748 Garching, Germany}

\affil{$^3$Walther-Meißner-Institut, Bayerische Akademie der Wissenschaften, 85748 Garching, Germany}

\affil{$^4$Munich Center for Quantum Science and Technology (MCQST), 80799 Munich, Germany}

\affil{$^*$Author to whom any correspondence should be addressed.}

\email{pablo.martinez-azcona@tum.de}



\begin{abstract}
Universal quantum computers require entanglement and non-stabilizerness, a resource known as \textit{quantum magic}. 
Here, we introduce a protocol that prepares magic steady states by leveraging non-Hermitian dynamics, which, contrary to unitary dynamics, can host pure-state attractors. By studying the dissipative qubit, we find the optimal parameters to prepare $\ket{H}$ and $\ket{T}$ steady states. Interestingly, this approach does not require knowledge or preparation of a particular initial state, since all the states of the Bloch sphere converge to the engineered target steady state. We also consider the addition of classical noise in the anti-hermitian part and provide the regimes for which the noisy dynamics still converges to high magic states. We also introduce a dissipative protocol to prepare magic steady states, compare the approaches with magic state cultivation and provide a particular realization of the non-Hermitian scheme in a cat qubit. 
\end{abstract}


\section{Introduction}
%
Clifford gates and Pauli measurements can be simulated efficiently on a classical computer using the stabilizer framework \cite{gottesman1998heisenberg, gottesman1997stabilizer, aaronson2004improved}. 
{Going  beyond the classical framework,} universal quantum computation {requires both} entanglement and \textit{non-stabilizerness}, also known as \textit{quantum magic} \cite{veitch2014resource}. {The latter} can be implemented through {the use of }non-Clifford $T$ gates or magic states \cite{bravyi2005universal}. {While }stabilizer codes \cite{gottesman1997stabilizer} are a cornerstone of quantum error correction \cite{lidar_quantum_2013}, non-stabilizerness is a key resource for fault-tolerant quantum computing. This resource can be quantified through several monotones such as the \textit{robustness of magic} \cite{howard2017application}, Wigner negativity \cite{howard2014contextuality}, or the \textit{stabilizer Rényi entropy} (SRE)  \cite{leone2022stabilizer, Oliviero2022, leone_Stabilizer_monotone_24}. High fidelity magic states are of paramount importance for quantum computing: they can be distilled from noisy ancillas through \textit{magic state distillation} \cite{bravyi2005universal}, a protocol which has been refined theoretically  \cite{bravyi2012magic, meier2012magic, haah2017magic} and experimentally realized in neutral atoms \cite{rodriguez2024experimental}. Recently, another method, known as \textit{magic state cultivation}, has been proposed. It leverages non-Clifford measurements to prepare magic states more efficiently \cite{gidney_magic_2024}. This protocol, which vastly reduces the number of physical qubits required to run Shor's algorithm at cryptographically relevant scales \cite{gidney_how_2025}, was realized experimentally in superconducting circuits \cite{rosenfeld_magic_2025}. Beyond quantum computing, the interest in non-stabilizerness extends to quantum foundations particularly through non-contextuality \cite{cusumano2025nonstabilizernessviolationschshinequalities, howard2014contextuality}, many-body systems \cite{Turkeshi2025, sierant2025fermionicmagicresourcesquantum, PhysRevB.111.054301, bera2025nonstabilizernesssachdevyekitaevmodel, jasser2025stabilizerentropyentanglementcomplexity, falcao2025magicdynamicsmanybodylocalized}, conformal field theory \cite{PhysRevB.103.075145}, molecular bonding \cite{sarkis2025moleculesmagicalnonstabilizernessmolecular}, measurement-induced phase transitions \cite{niroula2024phase}, as well as open quantum systems \cite{sticlet2025nonstabilizernessopenxxzspin, trigueros2025nonstabilizernesserrorresiliencenoisy, mittal2025quantummagicdiscretetimequantum}.

The presence of an environment critically affects the evolution of an open quantum system \cite{Breuer2007, Rivas2012}, which typically loses its quantum properties through decoherence \cite{zurek_decoherence_2003}. However, engineered dissipation can help prepare target steady states, opening a path for dissipative quantum computation \cite{verstraete2009quantum, Lin_2013, kraus_preparation_2008}.
The Lindblad master equation is often used to model a generic open quantum system \cite{Wiseman2009}. It can stem from various microscopic representations, and we will focus on two descriptions: 
Effective \textit{non-Hermitian} (NH) Hamiltonians \cite{bender1998real, Mostafazadeh2002}, which can be obtained by removing the quantum jumps through post-selection 
\cite{Wiseman2009, Jacobs2014, wang2019non}, as experimentally realized in superconducting qubits \cite{naghiloo2019quantum} and trapped ions \cite{quinn2023observing}; and \textit{stochastic Hamiltonians}, which provide a unitary but noisy model of certain master equations \cite{PhysRevA.64.052110, Kiely2021}, of relevance {for} quantum simulation \cite{PhysRevLett.118.140403} and information scrambling \cite{martinez_SOV_2023}. Combining these two approaches through stochastic perturbations in the anti-Hermitian part of the Hamiltonian leads to the \textit{antidephasing} master equation \cite{martinez_PRL25}.

Non-Hermitian Hamiltonians display several features not present in their Hermitian counterparts, such as exceptional points \cite{abbasi2022topological, wiersig_review_2020, miri_exceptional_2019}. These features can be used for certain quantum information tasks, such as {preparing a target steady state \cite{martinez_PRL25}}, speeding up entanglement generation \cite{Li_speeding_23}, or for a non-Hermitian version of the boson sampling problem  \cite{mochizuki_distinguishability_23}.  More generally, the computational complexity of generic NH evolution has been studied in \cite{barch2025computationalcomplexitynonhermitianquantum}.

In this article, we leverage non-Hermitian dynamics to prepare magic steady states in the single qubit example, which can be subject to noise in its decay rate.  
The paper is structured as follows: In Sec. \ref{sec:introMagic} we introduce the notion of quantum non-stabilizerness/magic and different measures for it, discussing the properties and relevance of each of these measures. In Sec. \ref{sec:diss_Qub}, we study the magic of the steady states in the Dissipative Qubit for different configurations.  We find the optimal parameters to prepare a maximally magic state, either a $\ket{H}$ or a $\ket{T}$ state, and calculate the evolution of the stabilizer Renyi entropy.
Section \ref{sec:SDQ} investigates the stability of the magic steady states under anti-Hermitian noise, particularly in the context of the \textit{stochastic dissipative qubit}. 
We numerically compute the robustness of magic of the steady state, and find that the non-Hermitian magic generation scheme is robust to the addition of anti-Hermitian noise \cite{martinez_PRL25}. We provide an analytical expression for the magic witness and discuss the interplay between the magic of the steady state and the speed at which we converge to it. 
Section \ref{sec:dissip_prot} discusses a protocol which prepares magic steady states from a pure Lindblad equation. This shows that our methods are not restricted to non-Hermitian dynamics, but they extend to other more general dissipative protocols.
In Sec. \ref{sec:discussion} we discuss the importance of the result in the context of the broader literature, beginning with a comparison (cf. Sec. \ref{sec:comparison}) with related approaches, such as magic state cultivation. Section \ref{sec:logical} discusses ways in which these protocols could extend to logical qubits, and proposes a particular protocol in the context of the cat qubit. Section \ref{sec:qjumps_vs_magic} shows why emission and absorption hinder non-stabilizerness, thus providing a motivation for post selection. Section \ref{sec:conclusion} provides some conclusion on our results and discusses future work.
Our results demonstrate a concrete pathway to \textit{non-Hermitian} and dissipative magic steady state generation, turning environmental decoherence into an asset for possibly fault-tolerant architectures.

\section{Quantum Non-stabilizerness, aka Quantum Magic}
\label{sec:introMagic}
It has been shown \cite{howard2014contextuality, bravyi2005universal} that entanglement by itself does not fully capture a quantum state’s computational power---in other words, its `quantumness'. Indeed, there are states that, despite exhibiting large amounts of entanglement, remain amenable to efficient classical simulation. In linear optics, this is the case of Gaussian states. For qubit-based systems, this happens for
stabilizer states, which can be computed through Clifford operations, namely elements of the Pauli group’s normalizer. 
\begin{itemize}
\item \emph{Stabilizer states:}
A state $\ket{\psi}$ is `\textit{stabilized}' by a unitary $\hat U$ if it is its eigenstate with eigenvalue  $+1$, i.e. $\hat U \ket{\psi}=\ket{\psi}$. 
We denote a generic Pauli string on $L$ sites as 
    $\hat\sigma_{\bs{\mu}} := \bigotimes_{\ell=1}^{L} \hat\sigma_{\mu_\ell}$, $\forall\bs{\mu}\in\{0,x,y,z\}^{\otimes L}$, where $\hat\sigma_0=\Id$. Consider the Pauli group as the closed group of Pauli strings  $\mathbb G_L =\{ i^k\hat \sigma_{\boldsymbol \mu}|\forall k \in \{0, 1, 2, 3\},\,  \boldsymbol \mu \in \{0,x,y,z\}^{\otimes L}\}$. The set of pure stabilizer states $\mathbb S_L$ are those states which are stabilized by a subgroup $\mathbb g_L \subset \mathbb G_L $ of $2^L$ elements of the Pauli group $\mathbb S_L = \{\hat s = \ket{\psi} \bra{\psi}| \hat U \ket{\psi} = \ket{\psi}, \hat U \in \mathbb g_L \subset \mathbb G_L, |\mathbb g_L|=2^L \}$\cite{aaronson2004improved}. 
    
    In the single qubit example, the set of stabilizer states is formed by the projectors on the eigenstates of the Pauli matrices, $\mathbb S_1 = \{\ket{0}\bra{0}, \ket{1}\bra{1}, \ket{\pm}\bra{\pm}, \ket{\pm i}\bra{\pm i}\}$, which set the vertices of the stabilizer octahedron, see Fig. \ref{fig:H-Tstate}(a,b). Prototypical instances of highly non-stabilizer states are the $\ket{H}=\cos\left(\tfrac{\pi}{8}\right)|0\rangle + \sin\left(\tfrac{\pi}{8}\right)|1\rangle$ and $\ket{T} =\cos \beta \ket{0} + e^{i \pi/4}\sin \beta\ket{1}$ states \cite{bravyi2005universal}, 
 the  $\beta$ angle being determined by $\cos(2 \beta) = \frac{1}{\sqrt{3}}$. These states are shown in Fig.~\ref{fig:H-Tstate}(a;b), together with their associated Clifford orbits.

    \item \emph{Non-stabilizerness:}
Since Clifford gates can be simulated in polynomial time on a classical computer via the Gottesman–Knill theorem \cite{gottesman1997stabilizer, gottesman1998theory, aaronson2004improved}, entanglement alone cannot guarantee a quantum speedup \cite{bravyi2005universal, howard2014contextuality}.
To transcend the limitations of stabilizer dynamics, one requires an additional resource commonly called magic or non-stabilizerness, as formalized in the resource theory of stabilizer computation \cite{bravyi2005universal, bravyi2012magic, howard2014contextuality}. Within this framework, Clifford gates and stabilizer states are deemed `free' (since they incur no overhead beyond classical simulability), whereas non-stabilizer states supply the `quantum overhead' necessary to outperform classical algorithms. 
%
%

 \item   { \emph{Measuring a resource:}} In the next sections, we {briefly} review different measures of the non-stabilizerness of a quantum state. This is a particular instance of a \textit{quantum resource theory} \cite{chitambar_QRT_19}.  
A resource theory specifies a set of states which are \textit{free} $\mathscr F$; in our case{, these are} stabilizer states. The notion of free states naturally induces a set of free operations $\mathscr O_\textsc{f}$ which map free states to free states, i.e. $\mathcal E \in \mathscr O_\ts{f}$ iff $\mc E(\hat \rho_\ts{f}) = \hat \rho'_\ts{f}$ and  $\hat \rho_\ts{f}, \hat \rho_\ts{f}' \in \mathscr F$. Apart from the free set, it is of key importance to quantify how much resource there is in a given non-free state. For this, we can introduce a \textit{measure} $M(\hat \rho)$ of the resource. A measure $M(\hat \rho)$ of a given resource is said to be \textit{faithful} if it vanishes for all free states $\ha \rho \in \mathscr F \Leftrightarrow M(\hat \rho)=0$. A measure is a \textit{monotone} if its value does not increase under a generic free quantum channel $M(\hat \rho)\geq M(\mathcal E(\hat \rho)), \, \forall \mathcal E \in \mathscr O_\ts{f}$. Lastly, we say that a measure is a \textit{pure state monotone} if it is a monotone but only for all the free channels which preserve pure states pure, i.e. for $\mathcal E_\textsc{p}(\ket{\varphi}\bra{\varphi})= \ket{\chi}\bra{\chi}$, the measure fullfills $M(\hat \rho)\geq M(\mathcal E_\ts{p}(\hat \rho))$.

\end{itemize}

\subsection{Stabilizer Renyi Entropy: A pure state monotone}

The stabilizer Rényi entropy (SRE) \cite{leone2022stabilizer, haug2023stabilizer} is a computable measure of quantum magic. It quantifies the spread of a quantum state over the Pauli basis. More specifically, for any $L$-qubit operator decomposed onto the Pauli basis as $\hat\rho = \frac{1}{2^L}\sum_{{\bs{\mu}}\in\{0,x,y,z\}^L}  \text{Tr}\left(\hat\sigma_{\bs{\mu}} \hat\rho\right)\,\hat\sigma_{\bs{\mu}}$,
the stabilizer Rényi entropy (SRE) is defined as
\begin{equation}\label{eq:SRE}
    M_\alpha(\hat\rho) := \frac{1}{1-\alpha}\log_2\left[\frac{1}{2^L}\sum_{\bs{\mu}\in\{0,x,y,z\}^L}\left|\text{Tr}\left(\hat\sigma_{\bs{\mu}} \hat\rho\right)\right|^{2\alpha}\right]\,.
\end{equation}
The SRE is a faithful quantity for pure states \cite{leone2022stabilizer}.
For $\alpha\geq 2$ it is a pure state magic monotone \cite{leone_Stabilizer_monotone_24, haug2023stabilizer}, i.e. it is non-increasing under free quantum channels that map pure states to pure states. For the maximally magic single qubit states, the SRE is $M_2(\ket{H}) = \log_2(4/3)$ and $M_2(\ket{T}) = \log_2 (3/2)$.

\subsection{Robustness of Magic (RoM)}
The SRE is only a monotone for pure states, and in general, a free quantum channel can make a pure state mixed. We now introduce a magic monotone that applies to both pure and mixed states but requires further optimization, thereby limiting its applicability to many-body systems. The \textit{Robustness of Magic} (RoM) \cite{howard2017application, heinrich2019robustness, seddon_quantifying_2019} is a faithful magic monotone for a general state $\hat \rho$, {which may be mixed}. It is inspired by similar quantities introduced to measure other resources, such as entanglement \cite{vidal_robustness_1999}. It can be defined from the decomposition of a given state $\hat \rho$ in the set of pure $L$-qubit stabilizer states $\mathbb S_L = \{\hat s_j\}_{j=1}^{|\mathbb S_L|}$ as 
\begin{equation}
    \mr R(\hat \rho) := \min_p \left\{\sum_j |p_j|\bigg | \ha \rho = \sum_{j=1}^{|\mathbb S_L|}p_j \ha s_j\right\},
\end{equation}
where $|\mathbb S_L|= 2^L \prod_{k=0}^{L-1}(2^{L-k} + 1) \sim 2^{L^2}$ is the size of the set of pure stabilizer states on $L$ qubits \cite{aaronson2004improved}. Intuitively, this quantity measures the distance between the considered state and the stabilizer polytope, the latter having vertices on the pure stabilizer states. 
Note that normalization enforces $\sum_j p_j=1$, but in general, the $p_j$ can be negative and thus constitute a quasiprobability \cite{pashayan_estimating_2015}. 
For a stabilizer state, $p_j>0 \:\: \forall j$, and therefore $R(\hat \rho_{\rm stab}) = \sum_j |p_j|=\sum_j p_j = 1$. By contrast, a non-stabilizer state necessarily has some $p_j<0$ and thus $R(\hat \rho_{\rm non-stab})>1$.

The robustness quantity can be computed from the linear program \cite{howard2017application} 
\begin{equation}
    \mr R(\ha \rho) = \min_p \lVert \bs p\rVert_1, \text{ subject to } \bs S \bs p = \bs r,
\end{equation}
where $\lVert \bs p\rVert_1 = \sum_j |p_j|$ denotes the $L_1$ norm of $\bs p$, $r_j = \Tr(\hat{\sigma}_j \hat \rho)$ are the components of the \textit{Pauli vector}, where $1 \leq j \leq 4^L$ counts the different Pauli strings. The matrix with elements $(\bs S)_{ij} = \Tr(\hat{\sigma}_i \ha s_j)$ encodes the projections of the Pauli strings over the pure stabilizer states; it is of size $|\mathbb S_L|\times 4^L$. 

Since the robustness is given by a linear program, it can be computed for a small number of qubits $L \leq 5$, or $L \leq 8$ using column generation, with larger system sizes being at reach under certain approximations \cite{hamaguchi2024handbook}. 
We compute the Robustness of Magic using the \href{https://github.com/quantum-programming/RoM-handbook}{RoM-handbook} library \cite{hamaguchi2024handbook}. 
Furthermore, taking the logarithm gives the \textit{log-free Robustness of Magic} (LRoM) \cite{timsina_robustness_2025}  
\begin{equation}\label{eq:LR}
    \mr{LR}(\ha \rho) := \log_2 \mr R(\ha \rho),
\end{equation}
which vanishes for states in the stabilizer polytope, $\mr{LR}(\hat \rho_{\rm stab})=0$. For the maximally magic single-qubit states, the LRoM takes the values
\begin{equation}
    \mr{LR}(|H\rangle) = \log_2 \sqrt 2, \qquad   \qquad \mr{LR}(|T\rangle) = \log_2 \sqrt 3.
\end{equation}

\subsection{A Magic Witness for mixed states} 
%
It would be desirable to have a quantity that can be computed without requiring extra optimization like the SRE, but which applies for mixed states. In this line, Leone et al. \cite{leone2022stabilizer} proposed to subtract the contribution from the Renyi entropy $S_2(\hat\rho)=- \log_{2} (\Tr (\ha \rho^2))$ and defined the mixed state SRE as
$\tilde M_2(\hat\rho) = M_2(\hat\rho) - S_2(\hat\rho)$.
However, this quantity is neither faithful nor a magic monotone: it can be non-zero for mixed stabilizer states, and can increase under free stabilizer operations. Nonetheless one can define a slightly different quantity, 
\begin{equation}
    \mathcal W_2(\hat \rho) = M_2(\ha \rho) - 3 S_2(\ha \rho),
\end{equation}
that lower bounds the RoM as  $\mc W_2(\ha \rho)\leq  2\mr{LR}(\ha \rho)$ \cite{haug_efficient_2026}. This quantity $\mc W_2$ is therefore a `witness' of quantum magic with the following properties:  a positive value $\mc W_2(\ha \rho)>0$ implies that the state $\ha \rho$ is outside of the stabilizer polytope, its LRoM being necessarily positive. Note that the converse  $\mc W_2\leq 0$ still allows for the state to be outside of the stabilizer polytope. Furthermore, since it is a lower bound, a large value of the witness $\mc W_2$ implies an even larger value of the LRoM. 
In the case of a single qubit, the witness is given by
\begin{equation}
\label{eq:1qubit_SRE}
    \mathcal W_2(\hat\rho) = -\log_2\left[4\frac{1 + x^4 + y^4 + z^4}{(1 + x^2+ y^2+ z^2)^3}\right]\,.
\end{equation}
where we denote the Bloch coordinates $\{r_x, r_y, r_z\}$ as $\{x,y,z\}$ to ease the notation. 
Throughout the text we will encounter cases in which we have to compute the non-stabilizerness of a mixed state, specially in sections \ref{sec:SDQ} and \ref{sec:dissip_prot}. In such cases, we will compute the LRoM numerically due to its monotonicity, but we will use the witness to get analytical insight into the magic of the target state.

\section{Generating magic in a dissipative Qubit}
\label{sec:diss_Qub}
Non-Hermitian systems have unique features, one of which is to naturally have pure-state attractors. 
We will exploit this feature to generate a high-magic steady state. We study the dissipative qubit (DQ) since it is one of the few NH Hamiltonians which has been realized in a pure quantum mechanical setup, particularly in superconducting circuits \cite{naghiloo2019quantum} and trapped ions \cite{quinn2023observing}. In the  $\{\ket{f}, \ket{e}\}$ basis, its Hamiltonian reads
\begin{equation}
    \label{eq:Hqubit}
\hat{H} = J_x \hat{\sigma}_x + J_y \hat{\sigma}_y + (\delta - i \Gamma ) \ha L = \begin{pmatrix}
    0 & J e^{- i \phi}\\
    J e^{ i \phi} & \delta - i \Gamma 
\end{pmatrix},
\end{equation}
with $J e^{- i \phi} = J_x - i J_y$ and  $\ha L =\hat\sigma_-\hat\sigma_+= \ket{e} \bra{e}$ the projector over the first excited state. 
Solving $(\hat{H} - \varepsilon_\pm)\ket{\psi_\pm} =0$ gives the eigenvalues and the right eigenstates 
\begin{align}
    \varepsilon_\pm =
    \delta_\pm - i \Gamma_\pm,
    \qquad \qquad \ket{\psi_\pm} = \frac{1}{\sqrt{J^2 + |\varepsilon_\pm|^2}} \Big(\begin{matrix} J e^{- i \phi}\\ \varepsilon_\pm  \end{matrix} \Big),
\end{align}
{where we have introduced $\Omega/2=(\Omega_R - i \Omega_I)/2 \equiv \sqrt{J^2 +\big((\delta - i \Gamma)/2\big)^2}$ to evidence the complex nature of the  eigenenergies,   $\delta_\pm = \delta \pm \Omega_R$ being the detuning correcting by the real part of $\Omega$, and $\Gamma_\pm = \Gamma \pm \Omega_I$  the corrected decay rate. }

Let us quickly summarize the phenomenology displayed by the dissipative qubit. With no detuning $\delta=0$, the system has two clearly separated phases: the $\mathcal{PT}$ symmetric (or unbroken) phase, and the $\mathcal{PT}$-broken phase{, respectively for $\Omega_I =0$ or not}. The former is for parameters $J^2-\Gamma^2/4>0$, with real eigenvalues (up to a constant imaginary irrelevant shift) yielding a purely oscillatory evolution. In the latter,  $J^2-\Gamma^2/4<0$ and the eigenvalues become imaginary, the dynamics thus exponentially converging to the eigenstate with the largest imaginary part, $\ket{\psi_+}$. At $\Gamma/J = 2$, the two eigenvalues and their corresponding eigenvectors coalesce, forming an exceptional point (EP). Adding a detuning ($\delta \neq 0$) breaks the $\mc{PT}$ symmetry: the eigenvalues are always complex, their real part encoding oscillatory dynamics and their imaginary part encoding exponential decay. This naturally gives the steady state as the eigenstate with the largest imaginary part,  $\ket{\psi_+}$.

\begin{figure}[h]
    \centering
\includegraphics[width = .7\linewidth]{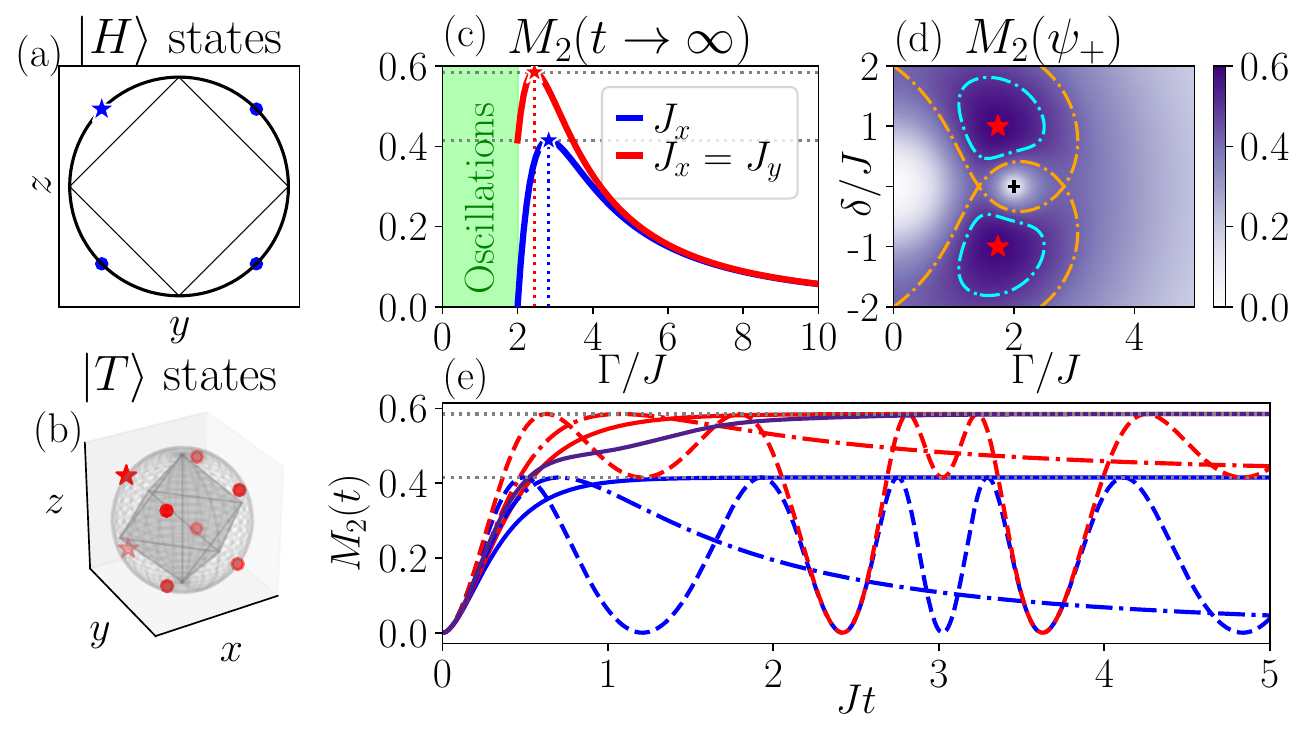}
    \caption{Bloch sphere representation of (a) $\ket{H}$ states ($x=0$) and (b) $\ket{T}$ states. The prepared states are represented with a star. The plots also show the Bloch sphere and the stabilizer octahedron (b), with their projections on the $x=0$ plane in (a). (c) Steady state SRE for cases \textit{(i)} Eqs. \eqref{eq:M2-Hstate} (blue) and \textit{(ii)} \eqref{eq:M2-Tstate} (red). SRE of the $\ket{H}$ and $\ket{T}$ states, $M_2(\ket{H})= - \log_2\frac{4}{3}\approx 0.415$ and $M_2(\ket{T})= - \log_2\frac{3}{2}\approx 0.585$ (dotted horizontal). (d) Steady State SRE as a function of the decay rate $\frac{\Gamma}{J}$ and the detuning $\frac{\delta}{J}$. We highlight the levels $M_2 = \log_2(4/3)$ (orange) and $M_2 = 0.5$ (cyan). A $\ket{T}$ state (red) appears for $\delta/J = \pm 1, \, \Gamma/J = \sqrt{3}$ and the EP $\delta = 0, \Gamma = 2 J$ (black cross). (e) SRE as a function of time for the $\mc{PT}$ broken phase (solid) at maximal magic point $\Gamma/J=2\sqrt{2}$ (solid blue) and $\Gamma/J= \sqrt{6}$ (solid red), for the Exceptional Point $\Gamma/J=2$ (dash dotted), and for the $\mc{PT}$ unbroken phase $\Gamma/J=1$ (dashed), for cases \textit{(i)} (blue) and \textit{(ii)} (red) of the DQ. The plot also shows case \textit{(iii)} (purple) at the maximum magic point $\Gamma/J = \sqrt{3}, \delta/J = 1$. The initial state is $\ket{\psi_0}=\ket{f}$.}
    \label{fig:H-Tstate}
\end{figure}

Now we characterize the non-stabilizerness of the steady state of the dynamics. Since it is a pure state, the Stabilizer Renyi entropy Eq.~\eqref{eq:SRE} is a good measure. It reads 
\begin{equation}
    M_2 (\ket{\psi_+})= -  \log_2 \left(\frac{1}{2} + \frac{(2J)^4\Big(\mr{Re}(\varepsilon_+^* e^{-i \phi})^4 + \mr{Im}(\varepsilon_+^* e^{-i \phi})^4\Big) + (J^2-|\varepsilon_+|^2)^4 }{2(J^2+|\varepsilon_+|^2)^4}\right)
    \label{eq:M2st_DQ}
\end{equation}
Let us look at particular cases where this expression greatly simplifies:
\begin{enumerate}[(i)]
    \item \underline{\textit{Zero detuning, real coupling:}} ($\delta=\phi=0$) In this case Eq.~\eqref{eq:M2st_DQ} reduces to
\begin{equation}\label{eq:M2-Hstate}
    M_2= -\log_2\left( 1 - \frac{4J^2(\Gamma^2-4 J^2)}{\Gamma^4} \right),
\end{equation}
which is only positive in the $\mc{PT}$ broken phase $\Gamma/J> 2$. Figure \ref{fig:H-Tstate}(c) shows the behavior of this expression (blue). At the exceptional point, $\Gamma = 2 J$, the SRE vanishes since the steady state is the stabilizer $\ket{-i}$ state. It then grows until it reaches a maximum of $M_2(\ket H) = \log_2(\frac{4}{3})$ at $\Gamma= 2 \sqrt{2} J $. At this point the $\ket{\psi_+}$ eigenstate corresponds to the $\ket{H}$ state with $y= - \frac{1}{\sqrt{2}}, \, z = +\frac{1}{\sqrt{2}}$---see Fig. \ref{fig:H-Tstate}(a).  In the large decay $\Gamma/J$ limit, the SRE decays quadratically with the decay rate $M_2 \sim 4 J^2  /( \ln(2) \Gamma^2)$. 
    
    \item \underline{\textit{Zero detuning, complex coupling:}}   
In the case $\delta = 0, \: \phi = \frac{\pi}{4}$, the SRE reduces to
\begin{equation}\label{eq:M2-Tstate}
    M_2= -\log_2\left( 1 - \frac{4J^2(\Gamma^2-3J^2)}{\Gamma^4} \right).
\end{equation}
This expression is shown in Fig. \ref{fig:H-Tstate}(c) (red). At the exceptional point, $\Gamma/J=2$, the SRE is that of an $\ket{H}$ state {(with $z=0$); as $\Gamma/J$ is increased}, it grows until reaching its maximum value of $M_2(\ket{T})=\log_2(\frac{3}{2})$ at $\Gamma=\sqrt{6} J$ and then decreases. This value of magic corresponds to $\ket{\psi_+}$ being a $\ket{T}$ state with maximal magic, namely $x = - y = z = \frac{1}{\sqrt{3}}$, see Fig. \ref{fig:H-Tstate}(b).  The large decay limit is the same as for the real coupling case mentioned above.
\item  \underline{\textit{Non-zero detuning }$\delta \neq 0, \, \phi = 0$}: 
In this case, we can prepare a $\ket{T}$ state with real coupling, making it easier for experimental implementation.  
The steady state SRE is detailed in the Appendix and shown in Fig. \ref{fig:H-Tstate}(d). 
The parameters yielding the maximum value of magic $M_2 = \log_2(3/2)$ are  $\delta = \pm J, \quad \Gamma = \sqrt{3} J
$, leading to the $\ket{T}$ state with Bloch coordinates  $\frac{1}{\sqrt{3}}=x = -y = \pm z$. Figure \ref{fig:H-Tstate}(d) shows the point of maximal magic, i.e. steady $\ket{T}$ state, and the regions where the steady state magic is higher than an $\ket{H}$ state $M_2\geq M_2(\ket{H})$ (orange), as well as the regions with $M_2 \geq 0.5$ (cyan). Note that on the line of zero detuning, $\delta/J=0$, we recover Eq.~\eqref{eq:M2-Hstate} for $\Gamma/J> 2$; in the $\mc{PT}$ unbroken regime, $\Gamma/J<2$, we can reach an $\ket{H}$ state. However, $\ket{\psi_+}$ is \textit{not} a steady state since, in this regime, all eigenvalues have identical imaginary parts and the dynamics does not converge to any particular state.  
\end{enumerate}



Figure \ref{fig:H-Tstate}(e) shows the time evolution of the SRE for the different cases in different regimes of parameters. Case \textit{(i)} (blue) shows that the magic oscillates in the $\mc{PT}$ unbroken phase (dashed line), grows and decays to zero at the EP $\Gamma/J=2$ (dash dotted) and grows saturating to the maximal value in the maximal magic point (solid line). The dynamics is restricted to $x=0$, bounding the magic by $M_2 \leq M_2(\ket{H}) = \log_2(4/3)$. Case \textit{(ii)} (red) shows similar behavior: oscillations in the $\mc{PT}$ unbroken phase (dashed), initial increase and then saturation to a $\ket{H}$ state at the EP (dash dotted) and saturation to a $\ket{T}$ state (solid). Case \textit{(iii)} (purple) saturates to the $\ket{T}$ state. 
In the $\mc{PT}$ unbroken phase, a state of maximal magic can be reached. But this requires stopping the evolution at the right time. The potential advantage of the non-Hermitian magic generation scheme is precisely the preparation of high-magic steady states.


\section{Generating magic in a stochastic dissipative Qubit}\label{sec:SDQ}
Having shown that highly magical states can be generated in a non-Hermitian setup, we now study the stability of these states under stochastic perturbations. We focus on perturbations on the antihermitian part, which account for fluctuations of the decay rate $\Gamma$.

\subsection{Antidephasing dynamics and steady states}
The effect of noise in the anti-Hermitian part of the Hamiltonian leads, for the noise-averaged density matrix, to an \textit{antidephasing} master equation that extends the set of possible steady states \cite{martinez_PRL25}. 
The \textit{stochastic dissipative qubit} (SDQ) can be described by the stochastic Hamiltonian 
\begin{align} \label{eq:hamsdq}
    \ha H^\ts{sdq}_t = J_x\ha \sigma_x + J_y \ha \sigma_y - i \big(\Gamma + \sqrt{2 \gamma} \xi_t \big)\ha L,
\end{align}
which is the dissipative qubit \eqref{eq:Hqubit} with Gaussian fluctuations in the decay parameter $\Gamma$. 
This noise, of strength\footnote{Contrary to our original work \cite{martinez_PRL25}, we now take $\gamma$ to have frequency dimensions so that the phase diagrams are easier to understand.} $\gamma$, is taken as a Wiener process $\mr d W_t  = \xi_t  \mr dt$ that follows It\=o rule $\mr d W^2_t = \mr d t$. 

This Hamiltonian generates an evolution $\ha U_{\mr d t} = e^{- i  \ha H^\ts{sdq}_t  \mr d t}$, updating the density matrix  $\uvrho_{\mr dt} = \ha U_{\mr d t}\ha \rho_0 \ha U_{\mr d t}\dg$. This density matrix is not normalized (as highlighted with the tilde $\ti \bullet$); and it depends on single realizations of the noise $\mr d W_t$. The normalized state of the system $\vrho_t$ follows the nonlinear stochastic master equation
\begin{align}\label{eq:SME}
    \mr d \vrho_t = (\uL[\vrho_t]- \av{\uL}_{\vrho_t}\vrho_t)\mr d t + (\ti{\mc M}[\vrho_t]-\av{\ti{\mc M}}_{\vrho_t}\vrho_t) \mr d W_t,
\end{align}
where the anti-dephasing Liouvillian and the measurement superoperator respectively read \cite{martinez_PRL25}
\begin{subequations}
\begin{align}
    \uL[\bullet]&=-i[\ha H_0, \bullet] - \Gamma \{\ha L, \bullet\} + \gamma \{\ha L, \{\ha L, \bullet\}\},  \label{eq:L}\\
    \ti{\mc M}[\bullet]&=-\sqrt{2\gamma}   \{\ha L, \bullet\},  \label{eq:M}
\end{align}
\end{subequations}
where $\ha H_0 = J_x \ha \sigma_x + J_y \ha \sigma_y$ is the qubit Hermitian Hamiltonian. We denote the expectation value of a superoperator as $\av{\ti{\mc X}}_{\ha \rho} \equiv \Tr(\ti{\mc X}[\ha \rho])$.

For each single trajectory, the density matrix of the SDQ, $\vrho_t = \frac{1}{2}(\Id + \bs{\st r}_t \cdot \ha {\bs \sigma})$, is governed by the stochastic master equation \eqref{eq:SME}. It is possible to find a set of coupled It\=o stochastic differential equations for the Bloch coordinates $\bs{\st r}_t=(\st x_t,\, \st y_t,\, \st z_t)$, namely 
\begin{equation}\label{eq:SDE_bloch}
      \begin{cases}
    \mr d \st x_t =\left(2 J_y \st z_t -(\gamma + \st z_t \tfrac{J B}{2})\st x_t\right) \mr d t - \sqrt{2 \gamma} \st x_t \st z_t \mr d W_t,  \\
     \mr d \st y_t = \left( - 2 J_x \st z_t -(\gamma + \st z_t \tfrac{J B}{2})\st y_t \right) \mr d t -  \sqrt{2 \gamma} \st y_t  \st z_t \mr d W_t, \\
     \mr d \st z_t =   \left(2(J_x \st y_t-J_y \st x_t) - \tfrac{J B}{2} (\st z_t^2 - 1) \right) \mr d t  - \sqrt{2 \gamma} (\st z_t^2 - 1) \mr d W_t,  
\end{cases}
\end{equation}
where  $J B= 2 \Gamma - 4 \gamma$ is a constant. We numerically integrate this system using the \textit{explicit order 1.5 strong scheme} from Kloeden and Platen \cite{Kloeden1992} (see Appendix).

Averaging over the noise leads to a closed master equation for the normalized density matrix  $\ha \rho_t= \frac{\avs{\uvrho_t}}{\Tr(\avs{\uvrho_t})}$ \cite{martinez_PRL25}
\begin{align}  \label{eq:MasterEq_SDQ}
    \partial_t \ha \rho_t =&- i [\ha H_0, \ha \rho_t] -(\Gamma -  \gamma)\{ \ha L, \ha \rho_t\} + 2 \gamma \Gamma^2\ha L \ha \rho_t \ha L \nonumber \\
    &+ (2 \Gamma - 4 \gamma) \Tr(\ha L \ha \rho_t) \ha \rho_t 
\end{align}
represented as $\ha \rho_t = \frac{1}{2}(\Id + \bs r_t \cdot \ha{\bs \sigma})$ on the Bloch sphere. {Averaging the coordinates \eqref{eq:SDE_bloch} yields }a set of coupled ordinary differential equations for the Bloch coordinates $\bs r_t=(x_t, y_t, z_t)$
\begin{equation}\label{eq:ODES_Bloch}
     \begin{cases}
    \dot x_t =\left(2 J_y z_t -(\gamma  + z \frac{J B}{2})x_t\right) ,\\
    \dot y_t = \left( - 2 J_x z_t -(\gamma  + z_t \frac{J B}{2}) y_t \right) , \\
    \dot z_t =   \left(2(J_x y_t-J_y x_t) - \frac{J B}{2} (z_t^2 - 1) \right) ,
\end{cases}
\end{equation}
which can be numerically integrated using a standard 4th order Runge-Kutta method.

The Liouvillian $\uL$ \eqref{eq:L} dictates the state evolution before renormalization, taken care with the last line of Eq.~\eqref{eq:MasterEq_SDQ}. We use the right eigendecomposition of this non-trace-preserving operator to find its eigenvalues $\ti{\mc L}[\ti \upsilon_\nu]= \lambda_\nu \ti \upsilon_\nu$. The steady state corresponds to the eigenvector $\ti \upsilon_0$ whose eigenvalue has the largest real part $\lambda_0$---assuming it is unique \cite{martinez_PRL25}; it is reached in a time governed by the inverse of the dissipative gap $\Delta = \min_{k>0}\mr{Re}(\lambda_0 - \lambda_k)$. The Bloch coordinates for the steady state in the $\phi=0$ case are (see Appendix)
\begin{equation}\label{eq:bloch_SDQx}
    x^\ts{s} = 0, \quad y^\ts{s} = \frac{2 \lambda_0 J}{4 J^2 + \lambda_0 (\lambda_0 - A J)}, \quad z^\ts{s} = -\frac{\lambda_0 (\lambda_0 - A J)}{4 J^2 + \lambda_0 (\lambda_0 - A J)},
\end{equation}
where $\lambda_0$ is the eigenvalue with the largest real part and $A J = \Gamma -\gamma$ is a constant. For $\phi=\frac{\pi}{4}$ they are
\begin{equation}\label{eq:bloch_SDQxy}
    x^\ts{s} = \frac{\sqrt{2}(B - 2 \lambda_0)}{B(\lambda_0-A)}, \quad y^\ts{s} = -\frac{\sqrt{2}(B - 2 \lambda_0)}{B(\lambda_0-A)}, \quad z^\ts{s}= \frac{(B - 2 \lambda_0)}{B}.
\end{equation}
\subsection{Real coupling ($\phi=0$)}\label{sec:real_hop_SDQ}

\begin{figure}[h]
    \centering
    \includegraphics[width=0.27\linewidth]{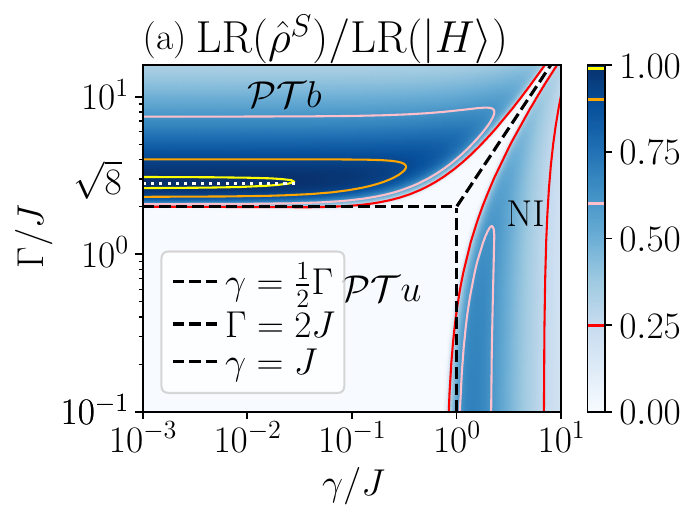}
    \includegraphics[width=0.27\linewidth]{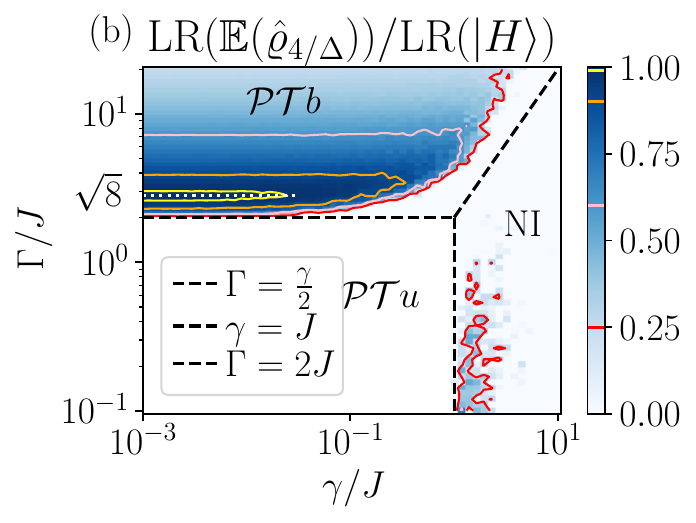}
    \includegraphics[width=0.44\linewidth]{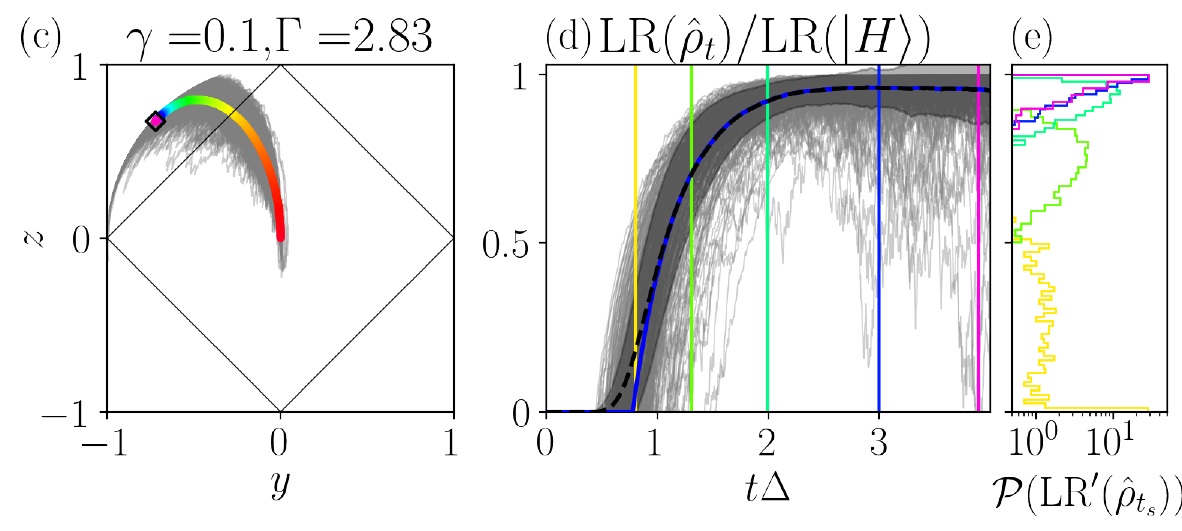}
    \caption{(a) Magic phase diagram of the analytical steady state \eqref{eq:SDQx_SRE_st}, (b) Magic phase diagram of the average state at $t=4/\Delta$. The time evolution has $N_t=150$ time-steps, averaged over $N_\mr{av}=100$ realizations. The initial state is fully mixed $\ha \rho_0 = \ha{\mbb 1}/2$. (c) Single trajectories in the Bloch sphere (gray lines), the average state (rainbow color line, time flows from red to blue). (d) Evolution of the relative LRoM of single trajectories (gray) as a function of time, along with: magic of the average state $\mr{LR}(\avs{\vrho_t})$ (blue), average of the magic of single trajectories $\avs{\mr{LR}(\vrho_t)}$ (dashed black) and its standard deviation. Time evolution has $N_t=1000$ time-steps, with $N_\mr{av}=1000$ trajectories [only $30\%$ are shown]. (e) Histograms of the SRE at different times (different colors).}
    \label{fig:SDQx_SRE}
\end{figure}

Figure \ref{fig:SDQx_SRE}(a) shows the phase diagram of the LRoM \eqref{eq:LR} computed for the analytical average steady state  \eqref{eq:bloch_SDQx} as a function of the decay rate $\Gamma$ and noise strength $\gamma$. The phase diagram highlights the three phases: $\mc{PT}$ unbroken, $\mc{PT}$ broken and Noise Induced (NI) \cite{martinez_PRL25}, the colorscale encodes the LRoM relative to the LRoM of a $\ket{H}$ state, the lines enclose the areas with more than $25\%, \, 60\%, \, 90\%$ and $99 \%$ (marked in the colorbar) of the $\ket{H}$ state magic.

The $\mc{PT}$b phase shows the highest value of the LRoM around the same point as the dissipative qubit $\Gamma = \sqrt{8}J$ (dotted white line). We see that the $\mc{PT}$b phase shows the highest value of the LRoM, which means that it is the most efficient to prepare a very magical steady state. The region with more than $99\%$ of the $\ket{H}$ state magic (yellow) survives until $\gamma /J \approx 0.03$, while the region with more than $90\%$ (orange) survives until $\gamma/J\approx 0.3$. The noise-induced phase shows relatively high magic, but not as much since its steady state is not as far from the stabilizer octahedron. In particular we see a region with more than $60\%$ of the $\ket{H}$ state magic around $\gamma/J\approx 1.5$, which survives until $\Gamma/J \approx 1.5$.

Figure \ref{fig:SDQx_SRE}(b) shows the phase diagram obtained from single realizations, as computed from the solution of the system of SDE's \eqref{eq:SDE_bloch} (with $\phi=0$) at a finite long time $\Delta \cdot  t_f = 4$---considered as representative of the steady state. The general behavior in the $\mc{PT}$b phase is quite similar to the average one, which corroborates the validity of expression \eqref{eq:SDQx_SRE_st}. The main differences arise in the large $\gamma/J$ limit, where the SDE integrator is more challenging, and the fact that some trajectories are inside the stabilizer polytope, thus pushing the average state closer to the octahedron and decreasing the magic of the average state. Still, we observe that the LRoM is non-zero in a  region similar to the $60\%$ one. 
We thus focus on the $\mc{PT}$b phase with small noise for the rest of the analysis. Figures \ref{fig:SDQx_SRE}(c-e) show the single trajectory simulations for $\gamma/J = 0.1$ at the maximal magic point $\Gamma/J = 2\sqrt{2}$, starting from the maximally mixed state $\ha \rho_0 = \frac{\Id}{2}$. The single trajectories (gray) in the cross section ($x=0$) of the Bloch sphere are seen to converge to the analytical steady state (black square). Same for the average trajectory $\avs{\vrho_t}$ (gradient thick line, the color represents time going from red to purple). The relative LRoM for single trajectories, illustrated in Fig.\ref{fig:SDQx_SRE}(d;e), is seen to be zero while the trajectories are inside the stabilizer polytope. After this time, the LRoM starts to grow. The plot also shows the evolution of the SRE for the average state $\mr{LR}(\avs{\vrho_t})$ (blue), the average of the single trajectory magic $\avs{\mr{LR}(\vrho_t)}$ (black dashed), and the standard deviation. The two averages do not coincide since the SRE is non-linear in the state, but we observe that $\avs{\mr{LR}(\vrho_t)}\geq \mr{LR}(\avs{\vrho_t})$ with their difference being very small, and only really notable at short times when some of the trajectories left the stabilizer octahedron but the average did not. After this initial regime, the two quantities show very similar values. Figure \ref{fig:SDQx_SRE}(e) shows the histograms of the relative LRoM, $\rm{LR}'(\hat \rho) = \rm{LR}(\hat \rho)/\rm{LR}(\ket{H})$, at different times. We observe a localization around the maximum value, and the logscale allows to see that the distribution is exponential around this value.   

\subsection{Complex coupling ($\phi=\pi/4)$}\label{sec:compHopSDQ}

\begin{figure}[h]
    \centering
    \includegraphics[width=0.27\linewidth]{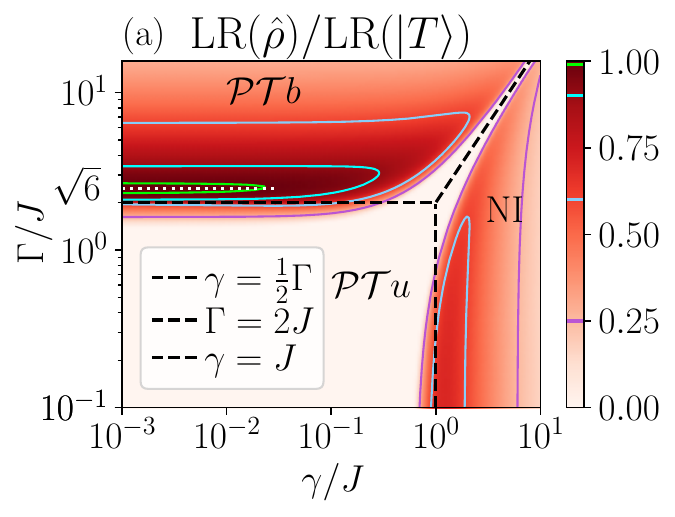}
    \includegraphics[width=0.275\linewidth]{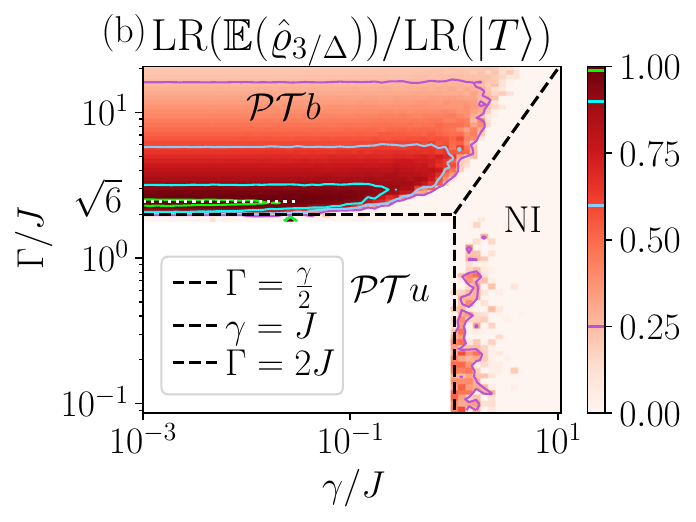}
    \includegraphics[width=0.44\linewidth]{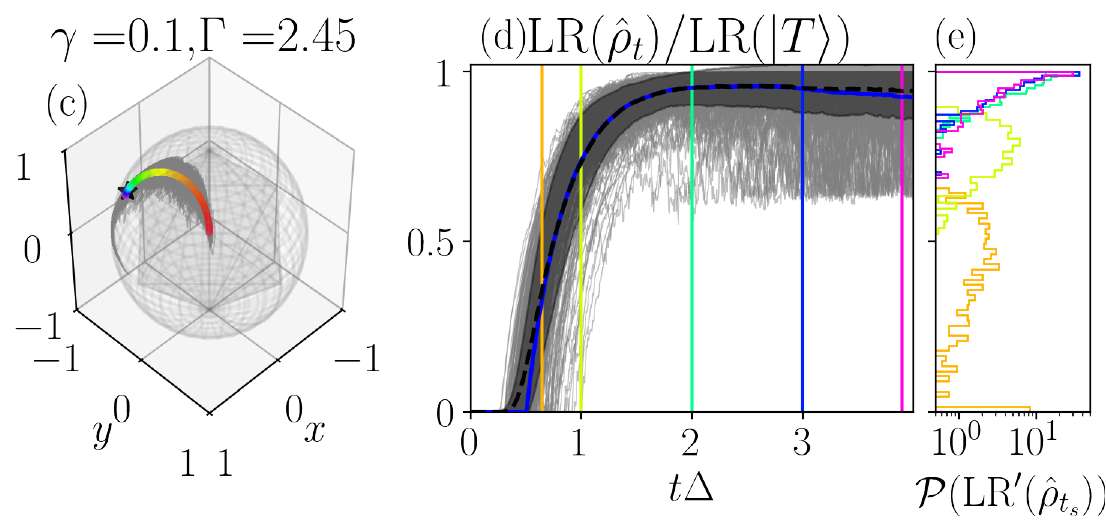}
    \caption{(a) Magic of the analytical steady state \eqref{eq:SDQxy_SREst}, (b) Magic phase diagram of the average state at $t=3/\Delta$. Time evolution has $N_t=400$ time-steps, averaged over $N_\mr{av}=100$ realizations. The initial state is fully mixed $\ha \rho_0 = \ha{\mbb 1}/2$. (c) Single trajectories in the Bloch sphere (gray lines), the average state (rainbow color line, time flows from red to blue). (d) Time evolution of SRE of single trajectories (gray), of the average state (blue line), and the average of single trajectory SRE (dashed black). Time evolution has $N_t=500$ time-steps, with $N_\mr{av}=1000$ realizations [only 200 shown]. (e) Histograms of the SRE at different times (different colors).}
    \label{fig:SDQxy_SRE}
\end{figure}

Figures~\ref{fig:SDQxy_SRE}(a;b) show the relative LRoM \eqref{eq:LR} phase diagrams for the steady states \eqref{eq:bloch_SDQxy} of the complex hopping case, which can prepare a $\ket{T}$ state. We again show the different phases, and highlight the levels corresponding to $25\%, \, 60\%, \, 90\%, \, 99\%$ of the $\ket{T}$ state LRoM. The plot shows that the region with the highest steady state non-stabilizerness corresponds to $\Gamma/J=\sqrt{6}$, with the steady state having more than $99\%$ LRoM of a $\ket{T}$ state up to $\gamma/J \approx 0.03$. We also see that at the exceptional point $\Gamma/J=2$ the LRoM is quite high, as seen previously in the noiseless case. The $\mc{PT}$b phase is the one with the highest amount of non-stabilizerness overall, with only very large decay rates $\Gamma/J\gtrsim 6$ having less than $60\%$ of the $\ket{T}$ state magic. The noise-induced phase also shows high values of the LRoM, but the maximum value of the LRoM is around $\max( \mr{LR}(\hat\rho_{\rm NI})) \approx 0.69 \mr{LR}(\ket{T})$.

Figure \ref{fig:SDQxy_SRE}(b) shows the relative LRoM from the average of the single trajectory simulations at a finite long time $t=3/\Delta$. We see a similar behavior in the $\mc{PT}$b phase, confirming that it is the most suitable to be used for the non-Hermitian magic preparation. At large noise we see a very notable difference arising from the same issue as before, numerical stability of the SDE solver and trajectories in the stabilizer octahedron pulling the state closer to the stabilizer set. However, there is one slight difference which persists at small noise: inside of the $\mc{PT}$u phase the dissipative gap is zero or extremely small, which means that we do not see a good convergence to a magical steady state.
Figure \ref{fig:SDQxy_SRE}(c) shows the trajectories in the Bloch sphere for the value of maximum steady state magic, preparing a $\ket{T}$ state from the maximally mixed state. 
Figures \ref{fig:SDQxy_SRE}(d;e) show the time evolution of the relative LRoM and the distributions for different times. Again, we see that the LRoM  starts to grow only after a certain initial time, reaching its maximum value in a time corresponding to two inverse dissipative gaps. The distributions in Fig. \ref{fig:SDQxy_SRE}(e) show that the LRoM concentrates around its maximum value of $\mr{LR}(\ket{T})$ and that the distributions is close to an exponential at long times.

\subsection{Analytical expression for Magic witnesses}
\label{sec:witness}

\begin{figure}[h]
    \centering
    \includegraphics[width=0.244\linewidth]{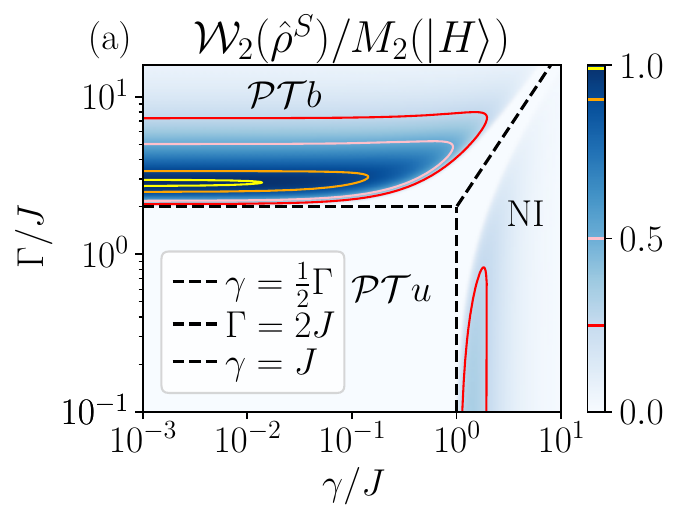}
\includegraphics[width=0.244\linewidth]{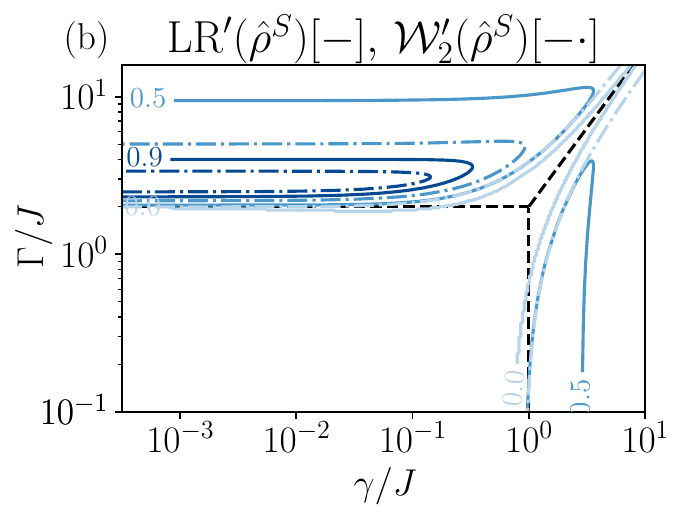}
\includegraphics[width=0.244\linewidth]{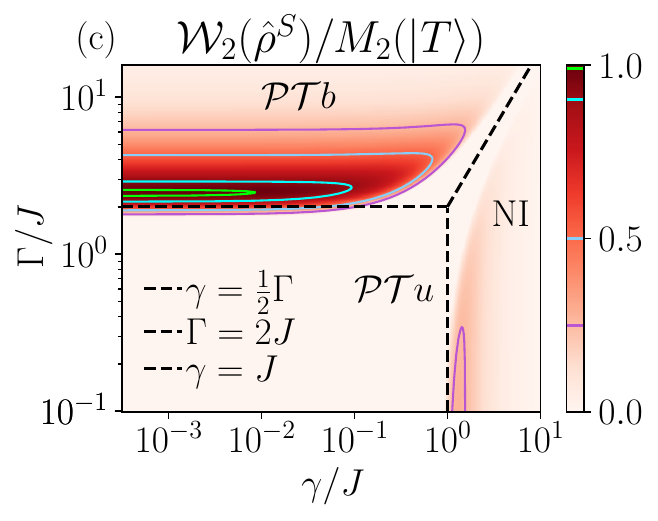}
\includegraphics[width=0.244\linewidth]{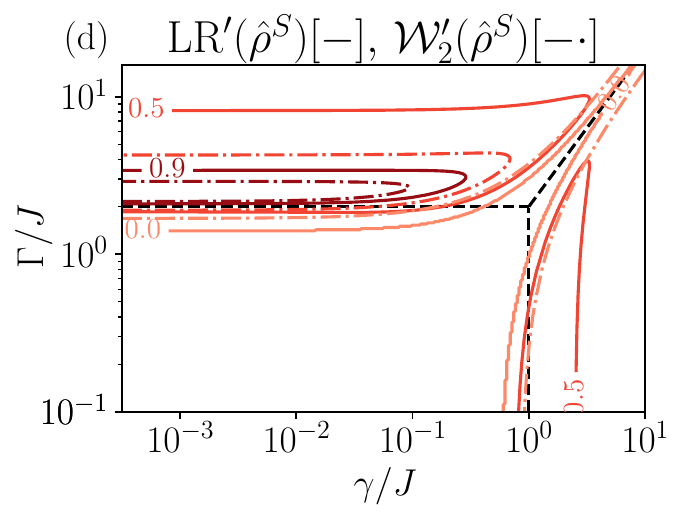}
    \caption{Magic Witnesses phase diagram for the steady state of the Stochastic Dissipative Qubit with: (a-b) real hopping and (c-d) complex hopping. (a, c) the phase diagram of the magic witness of the steady state \eqref{eq:SDQx_SRE_st}, \eqref{eq:SDQxy_SREst} relative to the maximum value that the witness can reach which is $M_2(\ket H)$ (a) and $M_2(\ket{T})$ (c). (b,d) show a comparison of the relative values of the witness with the relative values of the monotone, i.e. the LRoM, we highlight three levels $[10^{-10}, 0.5, 0.9]$. }
    \label{fig:SDQwitness}
\end{figure}
From the analytical form of the steady state $\hat \upsilon_0$ (see Appendix) we can obtain an analytical expression for the magic witness. 
For the real coupling case ($\phi= 0$), it reads 
\begin{align}\label{eq:SDQx_SRE_st}
    \mathcal W_2(\ha \rho^\ts{s})=-\log_2 \left(4\frac{1 + \lambda_0^4\frac{(\lambda_0-A J)^4 + 2^4 J^4}{(4J^2 + \lambda_0(\lambda_0 - A J))^4}}{(1 + \lambda_0^2\frac{(\lambda_0-A J)^2 + 4 J^2}{(4J^2 + \lambda_0(\lambda_0 - A J))^2})^3}\right).
\end{align}
 with the constant $A =\frac{\gamma - \Gamma}{J}$. Figure \ref{fig:SDQwitness}(a) shows this expression relative to its maximum value, $M_2(\ket{H})$. We see that the maximum value of the witness is obtained in the region around $\Gamma = \sqrt{8} J$ and that it decreases as we get away, or increase the strength of the noise. Note that the values of the witness for the NI phase are very small, with only a small region where the witness is more than $25\%$ of its maximum. We also compare the values of the relative LRoM (solid lines) and the relative witness (dash dotted lines), shown in Fig \ref{sec:witness}. The witness is non-zero in a region outside of the stabilizer polytope, but there are states outside of the polytope for which the witness is negative. This causes the regions where the LRoM is larger than a certain percentage of the maximum value to be more broad than those for the witnesss. This is particularly pronounces in the NI phase where the witness does not go over $50\%$ of the maximum value while the LRoM does. 

For the complex hopping case 
$\phi =\frac{\pi}{4}$, the steady state witness reads
\begin{equation}\label{eq:SDQxy_SREst}
    \mathcal W_2(\ha \rho^\ts{s}) = -\log_2 \left( 4 \frac{1 + \frac{(J B- 2 \lambda_0)^4}{B^4}\left(1 + \frac{8 J^4}{(\lambda_0 - A J)^4}\right)}{\left(1 + \frac{(J B- 2 \lambda_0)^2}{B^2}\left(1 + \frac{4 J^2}{(\lambda_0 - A J)^2}\right)\right)^3} \right),
\end{equation}
where we introduced the constant $JB = 2 \Gamma - 4\gamma$. This expression is shown in Fig. \ref{sec:witness} (c). We see a maximum value of the witness around $\Gamma/J = \sqrt{6}$ (green line), and a large value of the magic witness throughout the $\mc{PT}$ broken phase. Similarly to the previous case the values of the witness for the NI phase are small, with only a small region having more than $25\%$ of the SRE for a $\ket{T}$ state. Comparing the witness with the LRoM we see that similarly to the real hopping case, the areas with a certain percentage of the maximum value of the witness are narrower than those of the LRoM.

\subsection{Interplay between magic and speed of preparation}\label{sec:magic_speed}
\begin{figure}[h]
    \centering
        \includegraphics[width=0.3\linewidth]{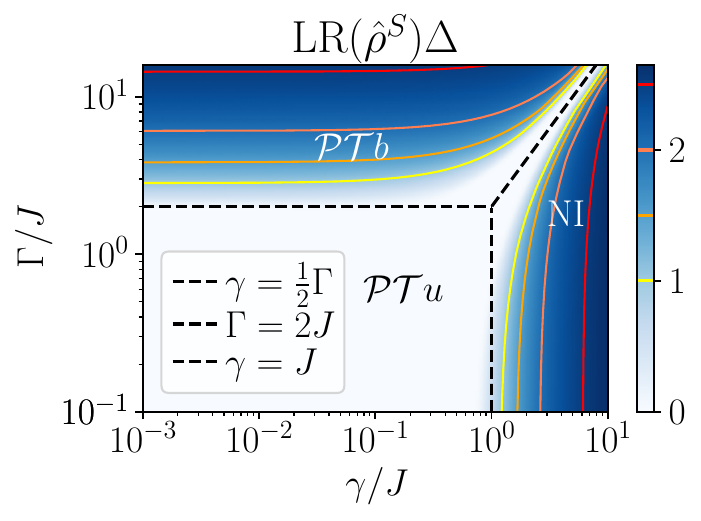}
        \includegraphics[width=0.3\linewidth]{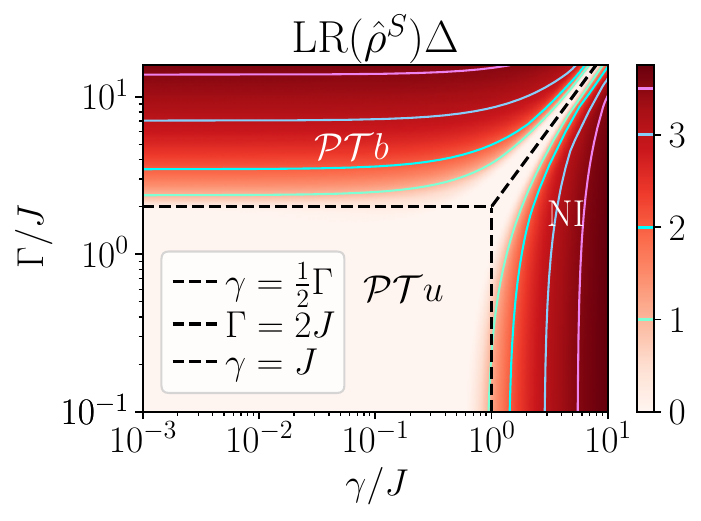}
        \caption{Phase diagram of the steady state LRoM times the dissipative gap for the SDQ with real (left) and complex (right) hopping.  The lines enclose the regions where the magic per unit time is above $\mr{LR}\Delta \geq 1, \, 1.5, \, 2, \, 2.5$ (left) and $\mr{LR}\Delta \geq 1, 2, 3, 3.5$ (right).}
    \label{fig:SRExgap_phasediag}
\end{figure}

One interesting feature enabled by the addition of noise is to prepare states with high magic, albeit not maximal, at a faster rate. As mentionned, the speed at which we prepare the steady state is given by the dissipative gap $\Delta$. Figure \ref{fig:SRExgap_phasediag} shows the phase diagrams of the product of steady state LRoM and the dissipative gap. The main idea is that we can produce less magic at a higher speed, which may compensate for the decaying success rate (see Appendix). In each of the plots for the real hopping (left) and complex hopping (right), we highlight the regions where the product is higher than certain values $1, \, 1.5, \, 2, \, 2.5$ for real hopping and $1, \, 2, \, 3, \, 3.5$ for complex hopping. Interestingly, we see that the maximum value of the steady LRoM per unit time increases as we go deep in the $\mc{PT}$b and NI phases. This means that even if the state we prepare is not very magical because it is close to the stabilizer states $\ket{e}$ or $\ket{f}$, we can prepare that state very quickly and the LRoM per unit time is higher than in the cases where we prepare an $\ket{H}$ or a $\ket{T}$ state more slowly. Note that the $\mc{PT}$b and NI phases have similar values of the product $\mr{LR}(\hat \rho^S)\Delta$, since the dissipative gap in the NI phase is larger than in the $\mc{PT}$b phase \cite{martinez_PRL25}, which compensates the smaller magic of the NI steady state. Comparing both  cases we see that the magic per unit time is larger in the complex hopping case (right), since the $\ket{T}$ state has more magic than the $\ket{H}$ state. The dissipative is gap is the same in both cases since the two Liouvillians have the same spectrum.

\subsection{Why Post-selection?: Emission and absorption hinder non-stabilizerness}
\label{sec:qjumps_vs_magic}
In this section we consider a more general framework beyond a single qubit and discuss the effect of typical quantum jumps (associated to emission and absorption) on magic. 
Consider a generic system of $L$ qubits with density matrix $\ha \rho = \frac{1}{2^L}(\ha{\mbb 1} + \sum_{\bs{\mu}} c_{\bs{\mu}} \ha \sigma_{\bs{\mu}})$, expressed in the set of Pauli strings $\{\ha \sigma_{\bs{\mu}}\}$.  Considering that each qubit interacts locally with its environment, the system evolves according to the Gorini-Kossakowski-Sudarshan-Lindblad (GKSL) master equation
\begin{equation}
    \partial_t \ha \rho_t = -i [\ha H, \ha \rho_t] + \sum_{k=1}^L (\mc D_{\ha \sigma^-_k}[\ha \rho_t]+\mc D_{\ha \sigma^+_k}[\ha \rho_t]),
\end{equation}
where we consider two dissipators $\mc D_{\ha X}(\bullet) := \gamma_{\ha X}\left (\ha X \bullet \ha X\dg - \frac{1}{2}\{\ha X\dg \ha X, \bullet \}\right )$ describing emission, and absorption respectively. 
The dissipators are composed of a  \textit{jump term} $\ha X \bullet \ha X\dg $ and an anti-Hermitian term  $- \frac{1}{2}\{\ha X\dg \ha X, \bullet \}$. 
Non-stabilizerness requires the state to be spread over many different Pauli strings. However, the jump term associated to emission or absorption destroys many of these Pauli strings since 
$\ha \sigma_k^\pm \ha \sigma_k^{(x,y)}\ha \sigma_k^\mp=0$  and only acts non-trivially on $\ha \sigma_k^{z}$ and $\Id_k$. 
So jump terms drastically reduce the complexity of Pauli strings, only keeping those which only involve $\ha \sigma^z_k, \, \Id_k$, thus greatly reducing the non-stabilizerness of the state. 
This is one general reason why non-Hermitian evolution, which post-selects these jumps, can be advantageous for generating magical states. Furthermore, based on this arguments, we expect that even a post-selection with a finite efficiency, such as done in hybrid-Liouvillians \cite{Minganti2020} would generate states with higher non-stabilizerness, due to the destructive effect of emission and absorption jumps.  We thus expect that the advantage demonstrated in the single-qubit example holds in more general setups. 

\section{Lindblad Dissipative Magic state preparation protocols}\label{sec:dissip_prot}

In this section, we develop a basic dissipative protocol to prepare magic states as steady states of a Lindblad equation \cite{lindblad_generators_1976, gorini_completely_2008, Breuer2007, Rivas2012} which for a single noise channel reads
\begin{equation}
    \partial \hat \rho_t = \mc L[\hat \rho_t] = - i [\hat H, \hat \rho_t] + \gamma\left( \hat L \hat \rho_t \hat L^\dagger - \frac{1}{2}\{\hat L\dg \ha L, \ha \rho_t\}\right).
\end{equation}
The steady states of this equation obey $\mc L[\rho_S]=0$.
The conditions under which the Lindbladian has a pure steady state $\mc L[\ket{\psi^\star}\bra{\psi^\star}]=0$ were derived by Kraus \textit{et al.} \cite{kraus_preparation_2008}. In particular, the target state has to be: an
eigenstate of the Hamiltonian $\hat H\ket{\psi^\star}= E \ket{\psi^\star}$ and a dark state of the jump operators $\hat L_k \ket{\psi^\star} = 0$. Therefore, for $\ket{\psi^\star}$ to be a magic state, the Hamiltonian has to be non-Clifford.

We consider the dissipative part only, with a single jump operator. 
To prepare the state $\ket{H} = \cos \frac{\pi}{8}\ket 0 + \sin \frac{\pi}{8}\ket 1$, the jump operator need to be of the form
\begin{equation}
\hat L_{\ket{H}} = \left(\begin{array}{cc}
    -a\sin \frac{\pi}{8} & a\cos \frac{\pi}{8} \\
    -b\sin \frac{\pi}{8} & b\cos \frac{\pi}{8}
\end{array}\right).
\end{equation}
 Setting the coefficients as $a = 1$, $b=0$, it can be decomposed as $\hat L_{\ket{H}} = e^{-i \frac{\pi}{8}}\hat \sigma^- \,\hat{\sigma}_z \,\hat{\mathsf{S}} \,\hat{\mathsf{H}} \,\hat{\mathsf{T}}\,\hat{\mathsf{H}}\,\hat{\mathsf{S}}\dg \,\hat{\sigma}_z$, where $\hat{\mathsf H}$ is the Hadamard gate, $\hat{\mathsf{S}} = \mr{diag}(1, i)$ is the phase gate and $\hat{\mathsf{T}}= \mr{diag}(1, e^{i \pi/4})$ is the T gate. Note that even if the jump operator $\hat L$ is non-unitary, as captured by the $\hat \sigma^-$ term, its compilation in terms of native gates still requires a non-Clifford unitary with a T gate. Considering the case $\hat H = 0$. The eigenvalues of the Lindbladian are $\lambda = \{0, -\frac{\gamma}{2}, -\frac{\gamma}{2}, -\gamma\frac{2 + \sqrt{2}}{4}\}$, giving a  dissipative gap $\Delta = \gamma/2$.

To prepare the state  $\ket T = \cos \beta \ket{0} + e^{i \pi/4} \sin \beta \ket{1}$ the jump operator should be of the form 
\begin{equation}
    \hat L_{\ket{T}} = \left(\begin{array}{cc}
    -a\sin \beta & a e^{- i \frac{\pi}{4}}\cos \beta \\
    -b \sin \beta & b e^{- i \frac{\pi}{4}}\cos \beta
\end{array}\right).
\end{equation}
Choosing again $a=1$ and $b=0$ the jump operator can be compiled as $\hat L_{\ket{T}} = e^{-i \beta}\hat \sigma^-\,\hat{\mathsf{T}}\dg \,\hat{\mathsf{H}}\,\hat{\mathsf{R}}_{2\beta} \,\hat{\mathsf{H}} \,\hat{\mathsf{T}},$ where we introduced the extra rotation $\hat{\mathsf{R}}_\theta = \mr{diag}(1, e^{i \theta})$. Again we see that compiling the non-unitary jump operator $\ha L_T$ requires non-Clifford unitaries, in particular two $T$ gates and the rotation by an angle $2 \beta = \arccos(1/\sqrt{3})$. The dissipative gap is the same as in the $\ket{H}$ state preparation case $\Delta = \gamma/2$.

\begin{figure}[h]
    \centering
    \includegraphics[width=0.35\linewidth]{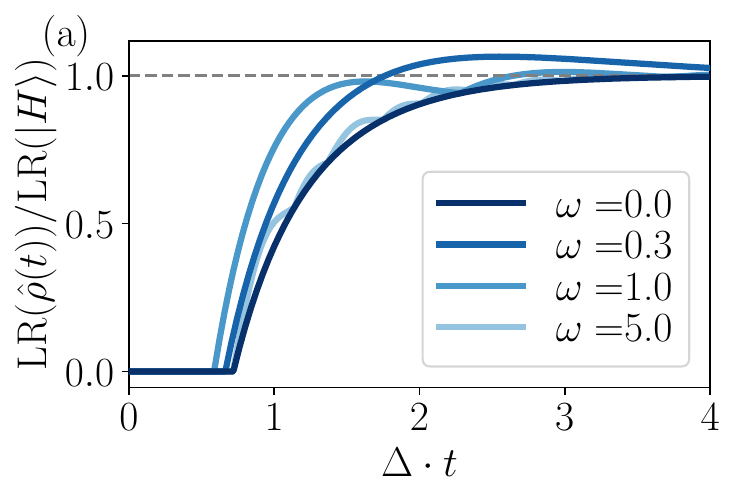}
\includegraphics[width=0.35\linewidth]{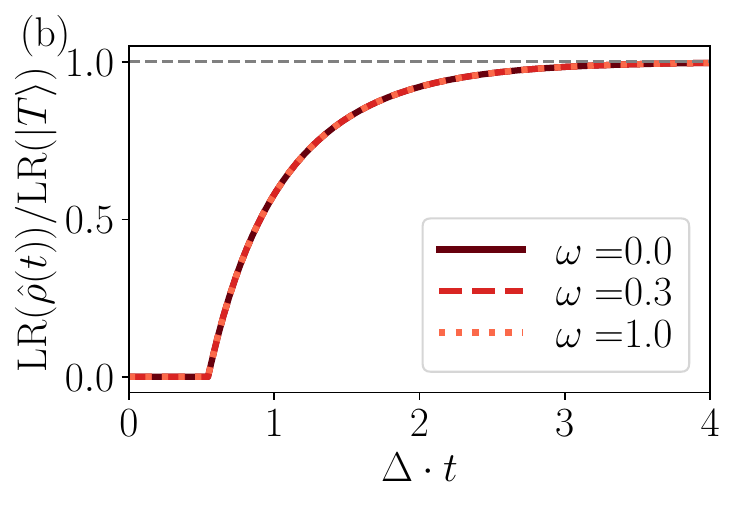}
    \caption{Evolution of the robustness of magic for the example Lindblad dissipative preparations of: (a) $\ket{H}$ states and (b) $\ket{T}$ states. Different colors and line-styles correspond to different values of $\omega$, while $\gamma=1$, but changing $\gamma$ only rescales the $x$ axis through the dissipative gap. The initial state is the maximally mixed state $\hat \rho_0 = \Id/2$.}
    \label{fig:dissip_prep}
\end{figure}

We illustrate the dissipative preparation scheme considering the following cases
\begin{equation}
    \begin{cases}
        \hat H_{\ket{H}} = \omega \ket{H}\bra{H}, \quad \hat L_{\ket{H}} = \begin{pmatrix}
            -\sin \frac{\pi}{8} & \cos \frac{\pi}{8}\\
            0&0\\
        \end{pmatrix}, \\
        \hat H_{\ket{T}} = \omega \ket{T}\bra{T}, \quad \hat L_{\ket{T}} = \begin{pmatrix}
            -\sin \beta & e^{-i \frac{\pi}{4}}\cos \beta\\
            0&0\\
        \end{pmatrix}, 
    \end{cases}
\end{equation}
and study the Robustness of magic of the time evolution generated by the corresponding Lindblad equations from an initial mixed state $\hat \rho_0 = \Id/2$. Figure \ref{fig:dissip_prep}(a) shows the $\ket{H}$-state preparation case for different values of $\omega$. When the frequency is zero $\omega=0$ we see that the LRoM starts at zero, while the state is inside the stabilizer octahedron, and then starts to increase monotonically towards the LRoM of an $\ket{H}$ state (reaching unity due to the normalization). When we progressively turn on the Hamiltonian, we see that the dynamics acquires some oscillations, but interestingly it never goes below the $\omega=0$ value. For small $\omega=0.3$ the oscillations are quite pronounced, and indeed bring us to a state with more LRoM than the $\ket{H}$ state, before saturating to $\ket{H}$. As $\omega$ increases we see that the period and the amplitude of the fluctuations decrease, giving an evolution close to the $\omega=0$ case for large $\omega=5$. Figure \ref{fig:dissip_prep}(b) shows the case where we prepare a $\ket{T}$ state. Interestingly, in this case, starting from the maximally mixed state shows the same evolution of the LRoM for all the values of $\omega$, starting at zero and then increasing monotonically towards the LRoM of a $\ket{T}$ state.

\section{Discussion}\label{sec:discussion}

Our method differs from Magic State Cultivation in two main ways: it does not require non-Clifford measurements and the magic state is prepared through the non-Hermitian system---generated by measurements, not through the collapse. Furthermore, post-selection is not central to our method, it is just the way in which NH Hamiltonians are implemented experimentally in quantum platforms at the moment. However, there are currently ideas to go beyond the need for post-selection: leveraging symplectic transformations of bosonic creation and annihilation operators \cite{Wakefield2024}, using Quantum Error Mitigation techniques \cite{kuji_quantum_2026}, or even through non-inertial frames \cite{Paiva2022}

\subsection{Comparison to other approaches}\label{sec:comparison}


In this section we compare our non-Hermitian and dissipative approaches with a method in the literature known as \textit{magic state cultivation} \cite{gidney_magic_2024, rosenfeld_magic_2025}. 


An alternative to the expensive magic state distillation protocols is \textit{magic state cultivation} \cite{gidney_magic_2024}, which prepares a magic state by performing fault-tolerant measurements of logical states and post-selects on the desired result. This protocol has recently been realized experimentally in superconducting circuits \cite{rosenfeld_magic_2025}, and promises to reduce the overhead in the number of physical qubits required for quantum advantage \cite{gidney_how_2025}.

The main idea is to inject a slightly magical state $\ha \rho_{\rm in}$ and perform measurements of the logical Hadamard operator $\hat H_L$ which is decomposed as $\hat H = \hat T_y \hat X \hat T_y\dg$, which involves a product of T gates along the $y$ direction $\hat T_y = e^{- i \frac{\pi}{8}\ha \sigma_y }$ and conditional measurements of $\hat X_L$ \cite{rosenfeld_magic_2025}. Then post-selecting on the measurement $+1$ on the ancilla yields the state\footnote{In here we follow the original notation by Bravyi and Kitaev \cite{bravyi2005universal} and refer to the magic states in which one of the Bloch coordinates is zero as $\ket{H}$ states, while the $\ket{T}$ states have all non-zero Bloch coordinates. The literature on Magic state cultivation \cite{gidney_magic_2024, rosenfeld_magic_2025} typically uses the naming $\ket{T}$ state also for a magic state with one null Bloch coordinate, which we refer to as an $\ket{H}$ state.} $\ket{\psi} = \ket{H}$.

\begin{table}[h]
    \centering
    \begin{tabular}{l|c|c|c|c|c|c|c|c|c|c}
         \textbf{Protocol}&  $\hat H_R$&  $\hat H_I$& $\Gamma/J$ & $\delta/J$ & $\phi$ & $\hat H$&  $\hat L$&  Meas.& Post-sel.?& State\\
         \hline
         NH:  Real&  Cl(P)&  Cl(P)& $\sqrt{8}$ & 0 &0 &  &  & & Yes & $\ket{H}$\\
         \hline
 NH: Complex& NCl(2P)& Cl(P)& $\sqrt{6}$ & 0 & $\frac{\pi}{4}$&  & & &Yes&$\ket{T}$\\
 \hline
  NH: Detuned & NCl(2P)& Cl(P)& $\sqrt{3}$ & 1 & 0 & & & &Yes &$\ket{T}$\
\\ \hline
         Dissipative Protocol &  & &&&  &  NCl&  NCl&  & No & $\ket{H}, \ket{T}$\\
         \hline
        Cultivation \cite{gidney_magic_2024, rosenfeld_magic_2025}&  & &&& &  &  &  NCl& Yes & $\ket{H}$\\
    \end{tabular}
    \caption{Comparison between the necessary ingredients for the different protocols to prepare magic steady states. We highlight whether each operator required is Clifford (Cl) or Non-Clifford (NCl), and whether the protocol requires post-selection or not. We also indicate for the NH protocols when the operator is a Pauli matrix (P), or a sum of two Pauli (2P) matrices. We stress out that even if the operators are Clifford the continuous time evolution generated by them will in general not be Clifford. }
    \label{tab:comparison}
\end{table}

Table \ref{tab:comparison} compares the approaches developed in this paper and Magic State Cultivation. The real hopping case $\phi=0$ generates a $\ket{H}$ state with a Clifford (Pauli) Hamiltonian and a Clifford (Pauli up to an irrelevant shift) anti-Hermitian part. In the other cases, either with complex hopping or with detuning, the Hermitian part of the Hamiltonian $\hat H_R$ is not Clifford, as it is given by a sum of two Pauli operators. In all the cases, however, the antihermitian part is a Pauli matrix up to the irrelevant shift. The table also specifies the optimal parameters to prepare a maximally magic state, and what that magic state is.

Interestingly, the dissipative protocol requires a non-Clifford Hamiltonian and a non-Clifford jump operator, necessitating $T$ gates and other particular non-Clifford rotations. This highlights the simplicity of the Non-Hermitian approach, which arises due to the conditions for the steady state of a NH Hamiltonian being less stringent than those of a pure state of a Lindbladian. The main strength of the dissipative protocol is that it does not require any post-selection, which means that the magic state is prepared deterministically and unconditionally.

In comparison, magic state cultivation requires both a non-Clifford measurement and post-selection to a particular measurement outcome. The need for post-selection is shared with NH Hamiltonians; however, the key difference is that since the post-selection in NH dynamics is done in real time, the success rate is exponentially decreasing in time. Within a time scale of one or two inverse dissipative gap, the post-selection success rate is relatively high (see Appendix). At longer times, this could become an issue but alternative proposals to implement NH dynamics without post-selection have been put forward. These include photonic gate decompositions of non-unitary operators  \cite{gao_photonic_2025}, symplectic transformations of creation annihilation operators \cite{Wakefield2024}, quantum error mitigation techniques \cite{kuji_quantum_2026} or even non-inertial frames \cite{Paiva2022}. 

The main advantage of magic state cultivation seems to be its direct implementation on a quantum error-corrected code, thus ensuring the preparation of logical magic states by design. In turn, the non-Hermitian and dissipative protocols discussed in this work apply to physical rather than logical qubits---mainly due to the current experimental platforms. The next section discusses how this gap could be bridged.



\subsection{Towards logical Non-Hermitian Magic steady states}
\label{sec:logical}
Focusing on the single non-Hermitian physical qubit, we have shown how the dynamics generated by non-Hermitian Hamiltonians, or by a specific Lindbladian, can be leveraged to prepare quantum states of importance for quantum information and quantum computing. 
However, non-Clifford operations---such as a $T$ gate or preparation of non-stabilizer states---are hard to implement on \textit{logical} qubits subject to Quantum Error Correction (QEC) \cite{lidar_quantum_2013}. 

In order to extend our framework preparing magic states to logical qubits, 
 we consider a \textit{proof-of-principle} realization of the non-Hermitian protocol in a cat qubit. This constitutes a physical architecture which is protected against certain types of errors. While not fully a logical qubit, this is a useful case to illustrates how our protocol may be applied to architectures requiring less overhead for fault-tolerance than physical qubits.

Autonomous QEC leverages the fact that the target state is an attractive steady state, thus cancelling errors around it \cite{xu_autonomous_2023, lachance-quirion_autonomous_2023, Li2024}. Our analysis of the stochastic anti-Hermitian perturbations in Sec. \ref{sec:SDQ} suggests that the target magic state can survive a certain level of stochastic perturbations. Let us detail a similar construction in a cat qubit, which can encode logical qubits \cite{cochrane_macrosc_99}. 
Specifically, we consider two-photon losses, with dissipator \cite{Mirrahimi2014,guillaud_catQ_23, Rglade2024}
\begin{equation}
    \partial_t \ha \rho = \kappa_2 \mathcal D_{\hat a^2 - \alpha^2}(\ha \rho).
\end{equation}
This dynamics includes the cat states $\ket{C^\pm} = \frac{\ket{+\alpha}\pm \ket{-\alpha}}{N_\pm}$, with $N_\pm=\sqrt{2(1\pm e^{-2 |\alpha|^2})}$ a normalization factor. Importantly, it stabilizes the code space $\mathscr H_c = \mr{span}(\ket{0_1}, \ket{1_c})$, the encoded qubit states reading 
\begin{subequations}
\begin{align}
    \ket{0_c} &= \frac{1}{\sqrt{2}}(\ket{C^+}+\ket{C^-}) = \ket{+\alpha} + \mc O(e^{-2|\alpha|^2}), \\
    \ket{1_c} &= \frac{1}{\sqrt{2}}(\ket{C^+}-\ket{C^-})= \ket{-\alpha} + \mc O(e^{-2|\alpha|^2}).
\end{align}
\end{subequations}
%
%
The main source of error for cat qubits is the logical $X$ phase flip, described by the single-photon loss channel $\kappa_1 \mathcal D_{\hat a}(\ha \rho)$. Note that under one application of the photon loss channel, the states in the code subspace remain in the code subspace; while higher order terms induce leakage of the code subspace \cite{dubovitski_bitflip_25}.

In the physical realization of the dissipative qubit \cite{naghiloo2019quantum}, the subspace $\{\ket{e}, \ket{f}\}$ is used to encode the qubit state, an additional level $\ket{g}$ being used to monitor leakage out of this subspace. To do something similar in the cat qubit, we introduce a jump operator which takes one of the encoded states out of the code subspace to a state which can be monitored, thus allowing us to determine whether a jump happened and do post-selection treatement. A possible operator is $\hat L_\perp = \ha a\dg(\ha a-\alpha)$, that acts on only one of the encoded states $\ket{1_c}$, giving it an additional photon such that it leaves the code subspace, $\hat L_\perp \ket{1_c}=\ket{\chi_\perp} \approx - 2\alpha \ha a\dg \ket{-\alpha} \notin \mathscr H_c$. This dissipator is similar in spirit to the one used to prepare stabilizer states \cite{kraus_preparation_2008, barreiro_open-system_2011}. Here, thanks to post-selection and to the anti-Hermitian part of the Hamiltonian being Clifford (cf. Sec. \ref{sec:comparison}), we can use a similar simple jump operator to generate a magic non-stabilizer steady state.  By monitoring the population of $\ket{\chi_\perp}$ and keeping only the trajectories where a jump did not occur, one induces a dynamics with an attractor. Thus, one can prepare  e.g. an $\ket{H}$ state. The full master equation for this reads
\begin{equation}
    \partial_t \hat \rho = -i J[\hat X_c, \hat \rho] +\kappa_2 \mc D_{\ha a^2 - \alpha^2}(\ha \rho) + \kappa_\perp \mathcal D_{\ha L_\perp}(\ha \rho),
\end{equation}
with a $\kappa_\perp$ decay rate on the dissipator $\mc D_{\ha L_\perp}= \mc D_{\hat a\dg(\ha a - \alpha)}$, and the $X$ Pauli matrix encoded in the logical qubits through $\ha X_c = \ket{0_c}\bra{1_c} + \mr{h.c.}$. Post-selecting on no jump to $\ket{\chi_\perp}$ trajectories, the dynamics remains in the code subspace. Then, when  $\kappa_2$  is large enough to be adiabatically eliminated, the projection of the density matrix on the code subspace $\ha \rho_c = \mathcal P_c(\ha \rho) = \sum_{j,k=0}^1 \ket{j_c}\braket{j_c|\ha \rho|k_c} \bra{k_c}$ evolves according to the equation
\begin{equation}
    \partial_t \ha \rho_c = -i J [\ha X_c, \ha \rho_c] - 2\kappa_\perp |\alpha|^2 (1+|\alpha|^2)\{\ket{1_c}\bra{1_c}, \ha \rho_c\},
\end{equation}
where the antihermitian term can be expressed in the code subspace as $\mc P_c(\ha L_\perp\dg \ha L_\perp )\approx  4|\alpha|^2 (1 + |\alpha|^2)\ket{1_c}\bra{1_c}$, 
which is precisely the studied non-Hermitian dynamics with decay rate $\Gamma = 2\kappa_\perp |\alpha|^2(1+|\alpha|^2)$. Therefore, the encoded $\ket{H_c}$ state can be prepared by setting $\Gamma = 2\kappa_\perp |\alpha|^2(1+|\alpha|^2)= \sqrt{8}J$, all under the assumption $\Gamma, \, J\ll \kappa_2$ to justify the adiabatic elimination. It is possible to prepare a $\ket{T}$ state with smaller decay rate. For this, we consider the detuned case with Hamiltonian $\delta \ha Z_c + J \ha X_c$, which follows the equation
\begin{equation}
    \partial_t \ha \rho_c = -i [(J \ha X_c+\delta \ha Z_c), \ha \rho_c] - 2\kappa_\perp |\alpha|^2 (1+|\alpha|^2)\{\ket{1_c}\bra{1_c}, \ha \rho_c\}.
\end{equation}
Choosing the parameters $\delta = \pm J$, and $ \Gamma = 2\kappa_\perp |\alpha|^2 (1+|\alpha|^2) = \sqrt{3} J$, and post-selecting on the quantum jumps, the evolution prepares a $\ket{T}$ state.


Finally, we stress out that the choice of jump operator is not unique, the only condition being that when projected on the code space, it should be of the form $\ha L_\perp = \ket{\chi_\perp}\bra{1_c}$ or $\ha J = \ket{\chi_\perp}\bra{0_c}$, where $\ket{\chi_\perp}\in \mathscr H_c^\perp$, so that:
\begin{enumerate}[(i)]
    \item the jump $\ket{1_c} \to \ket{\chi_\perp}$ can be detected,
    \item the anticommutator gives the desired antihermitian term inside of the code space $\ha L_\perp\dg \ha L_\perp \propto \ket{1_c}\bra{1_c}$ or $\ha L_\perp\dg \ha L_\perp \propto \ket{0_c}\bra{0_c}$.
\end{enumerate}


\section{Conclusions} \label{sec:conclusion}

This article shows how to exploit the dynamics generated by a non-Hermitian Hamiltonian for the analog preparation of magic steady states. Looking at the single qubit example, we found the optimal parameters to prepare the $\ket{H}$ and $\ket{T}$ states, considering the cases of: no detuning with eigher real or coupling, and non-zero detuning. We showed how the preparation of magical states survives the application of noise in the anti-Hermitian part of the non-Hermitian Hamiltonian, highlighting the noise strengths still allowing for a highly magical steady states. The noise-induced phase of the stochastic dissipative qubit \cite{martinez_PRL25} can also host magical steady states, although the non-stabilizerness is lower than in the $\mathcal{PT}$ broken phase. The dynamics at the level of single realizations of the noise showed that the trajectories can exponentially concentrate around the maximum value of the non-stabilizerness in the weak noise regime. We also proved that the non-stabilizerness of the steady state displays an interesting dependence on the speed of preparation: it is possible to prepare non-maximal magic steady states but at a faster rate than the maximally magic states. 

We consider a fully dissipative, GKSL-type protocol (no post-selection on the jumps) to prepare magic steady states. Interestingly, we find that the jump operators then require non-Clifford jump operators, also required for magic state cultivation. This restriction is not needed for the non-Hermitian approach. We proposed a proof of principle protocol that could be implemented on a cat qubit, which self corrects under certain types of errors. Lastly, we finished with a discussion on the effect of emission and absorption on the non-stabilizerness of a certain state, highlighting that we expect this idea to apply to systems beyond a single qubit, or with a finite efficiency of post-selection. 

A remarkable feature of the proposed approach over standard distillation or cultivation algorithms is that \textit{any} state, even the maximally mixed one, can be turned into a magical state---not only those with a certain amount of magic. This `magic-charged' qubit can then be interfaced via standard Clifford swaps or teleportation to inject the distilled magic into any target qubit. This modular approach supports parallel charging of ancillas and decouples high-overhead non-Clifford preparation from the main computation, while keeping all transfer steps within the efficient stabilizer framework.

\emph{Note added:} After completion of this work, we became aware of the related articles studying magic in non-Hermitian contexts \cite{Karmakar:2025jmg, tirrito2025magicphasetransitionsmonitored}, however, they focus on complementary issues to the ones discussed here, namely: a novel protocol to engineer NH Hamiltonians with applications to magic state distillation and magic phase transitions of monitored fermions, where the NH Hamiltonian arises in the no-click limit.

\emph{Acknowledgments}--- We thank Aritra Kundu, Komal Kumari, Myungshik Kim, Pablo Sala, Aashish Clerk, Poetri Sonya Tarabunga and Peter Rabl for insightful discussions. This research was partly funded by the Luxembourg National Research Fund (FNR, Attract grant QOMPET 15382998). The authors acknowledge funding via the FNR-CORE Grant ``BroadApp'' (C20/MS/ 14769845) and ERC-AdG Grant ``FITMOL'' (No. 1110643). P.M.A. also acknowledges
funding from the Swiss National Science Foundation (Project No. CRSII 222812/1). This research
is part of the Munich Quantum Valley initiative,
which is supported by the Bavarian state government with funds from the Hightech Agenda Bayern Plus.

\emph{Code and Data Availability}: The reader will find an open-source Python and Mathematica code to reproduce the results of this paper at \esrgithub.

\printbibliography

\newpage
\appendix


\section{Dissipative qubit}
\subsection{Success Rate}
A non-Hermitian Hamiltonian $\ha H_\ts{r} - i \ha H_\ts{i}$ can be generated, in a quantum mechanical setup, through the post-selection of quantum jumps. The probability of sampling a non-Hermitian trajectory, also called \textit{success rate}, is given by the norm of the \textit{unnormalized} state $\ket{\ti \psi_t} = e^{- i (\ha H_\ts{r} - i \ha H_\ts{i})t} \ket{\psi_0}$, as 
\begin{equation}\label{eq:27}
    \mr{SR}_t := \braket{\ti \psi_t|\ti \psi_t}.
\end{equation}

\begin{figure}[h]
    \centering
    \includegraphics[width=0.4\linewidth]{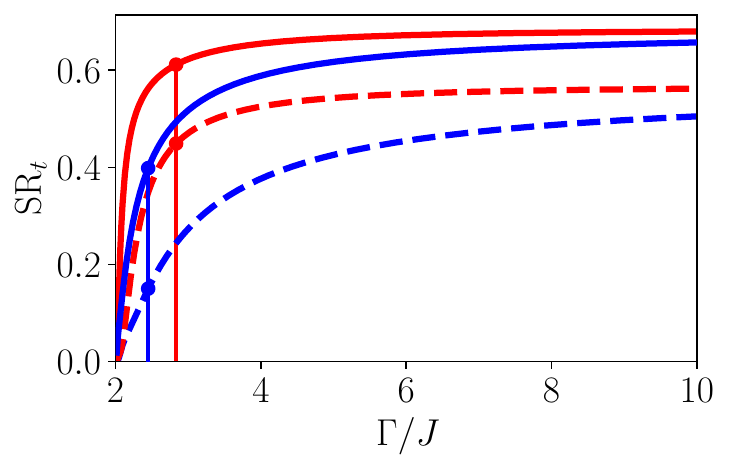}
    \caption{Success rate \eqref{eq:27} of the dissipative qubit for (i) $J_x=J$ (red) and (ii) $J = J_x - i J_y$ (blue) at $t= 1/\Delta$ (solid) and $t = 2/\Delta$ (dashed) as a function of $\Gamma/J$. The vertical line highlight the parameter  $\Gamma/J$ for which the steady state reaches the highest magic point, the success rate then being $\mr{SR}_{1/\Delta, \Gamma^*}^{(i)}=0.61,\, \mr{SR}_{2/\Delta, \Gamma^*}^{(i)}=0.45$ and $\mr{SR}_{1/\Delta, \Gamma^*}^{(ii)}= 0.41,\, \mr{SR}_{2/\Delta, \Gamma^*}^{(ii)}= 0.16$. The curves depend on the initial state, taken here to be $\ket{\psi_0} = \ket{+} = \frac{\ket{e}+\ket{f}}{\sqrt{2}}$}
    \label{fig:successRate}
\end{figure}

Figure \ref{fig:successRate} shows the success rate for the dissipative qubit initially in $\ket{\psi_0} = \ket{+}$ in the $\mc{PT}$ broken regime for the real (red) and complex (blue) coupling cases.  The values are given for a time-scale dictated by the dissipative gap $t=1/\Delta$ (solid line) and twice that time-scale (dashed line). The success rate decays to zero in the infinite time limit, for this reason, we want to stop the protocol as soon as we reach close to the target steady state, although our protocol prepares magic states from any state, carefully selecting an initial state could help mitigate the success rate decay. 

\subsection{Bloch coordinates and SRE for the steady state of the dissipative qubit}
The Bloch coordinates $r_i = \braket{\psi_+|\ha \sigma_i|\psi_+}$ of the steady state $\ket{\psi_\pm} = \frac{1}{\sqrt{J^2 + |\varepsilon_\pm|^2}} \Big(\begin{smallmatrix} J e^{- i \phi}\\ \varepsilon_\pm  \end{smallmatrix} \Big)$ of the Dissipative Qubit read
\begin{equation}
    x = \frac{2 \mr{Re}(\varepsilon_+^* J e^{- i \phi})}{J^2+|\varepsilon_+|^2}, \, y = -\frac{2 \mr{Im}(\varepsilon_+^* J e^{- i \phi})}{J^2+|\varepsilon_+|^2}, \, z = \frac{J^2-|\varepsilon_+|^2}{J^2+|\varepsilon_+|^2},
\end{equation}
from where the steady state SRE \eqref{eq:M2st_DQ} immediately follows. 


   
In the case $\phi=0$ and $\delta\neq 0$, the SRE of the steady state follows as

\begin{eqnarray}
    M_2 &=& - \log_2 \left( \frac{1}{2}(1+ x^4 + y^4 + z^4)\right)\nonumber \\
    &=&-\log_2 \left(
    \frac{1}{2}\left[1 + \left(\frac{\varepsilon_- + \varepsilon_+^*}{\varepsilon_- - \varepsilon_+^*} \right)^4 + 2^4 \left(\frac{\varepsilon_-}{\varepsilon_+}\right)^2\frac{\delta_+^4 + \Gamma_+^4}{(\varepsilon_- - \varepsilon_+^*)^4}\right]
    \right).
\end{eqnarray}




\section{Details on the Stochastic Dissipative Qubit}
In here we introduce some further details concerning the Stochastic Dissipative Qubit example of the main text. 
\subsection{The vector field in the Bloch sphere}
\begin{figure}[h]
    \centering
    \includegraphics[width=.7\linewidth]{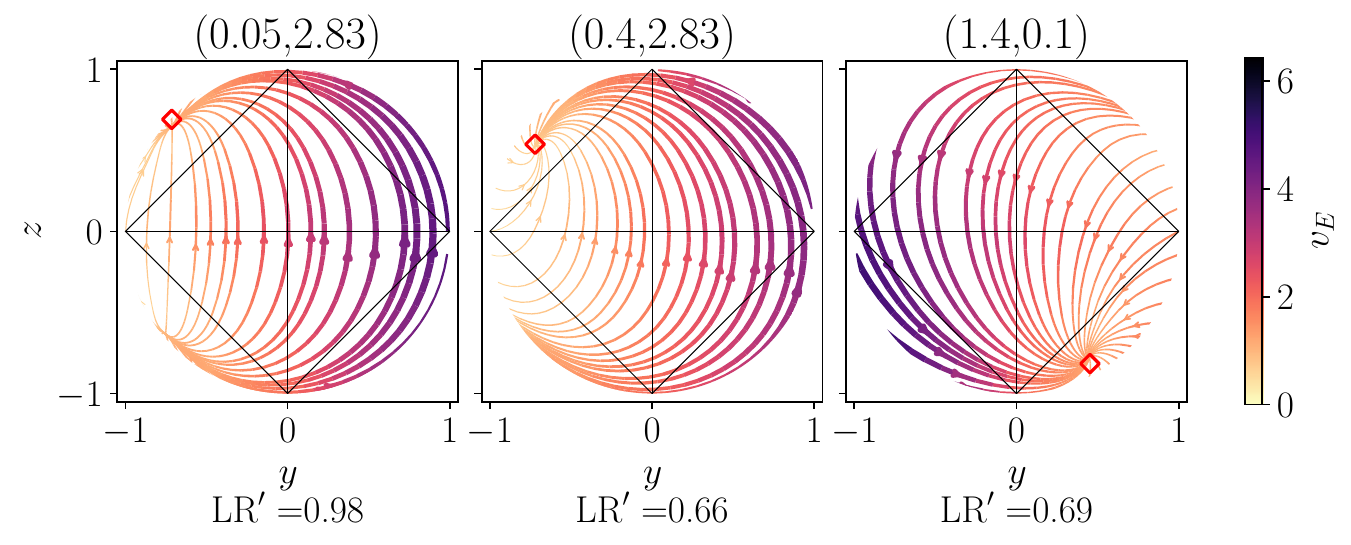}
    \caption{Streamlines of the vector field in the Bloch sphere for different parameters $(\gamma, \Gamma)$, corresponding to: (left) very weak noise and optimal decay rate, (center) intermediate noise and optimal decay rate, and (right) Noise Induced phase. The colorscale and thickness of the streamlines represent the Euclidean speed $v_E = \sqrt{(\dot y)^2+(\dot z)^2}$. The figure also states the LRoM relative to the $\ket{H}$ state.}
    \label{fig:SDQ_streamlines}
\end{figure}

Figure \ref{fig:SDQ_streamlines} shows the streamlines of the vector field in the Bloch sphere for the SDQ model with $J_y=0$. We see that for small values of the noise the dynamics converges to the surface of the Bloch sphere, in particular we converge close to the $\ket{H}$ state. 
\subsection{Real hopping SDQ}
In this section we recall the spectral analysis carried out in \cite{martinez_PRL25}. For the SDQ with real hopping the Liouvillian reads 
\begin{equation} \label{eqSM:L}
	\uL = J \left( \begin{array}{cccc}
		0 & i & - i & 0\\
		i & A & 0 & -i \\
		- i & 0 & A & i \\
		0 & - i & i & B
	\end{array}\right),
\end{equation}
where $J A = \gamma -\Gamma$ and $J B = 4 \gamma -2 \Gamma$. The spectrum is given in terms of the auxiliary functions
\begin{align}
    C &= \frac{4 + AB}{3} - \left( \frac{A+B}{3}\right)^2, \\
    D &= \frac{A + B}{3}(4 + AB) - 2 B - 2 \left(\frac{A+B}{3} \right)^2, \\
    U_{m, \pm} &= e^{i \frac{2 \pi}{3}m} \left( - \frac{D}{2} \pm \sqrt{\left( \frac{D}{2}\right)^2 + C^3} \right)^{1/3}, \\
    \Lambda_{m, \pm} &= J \left( U_{m, \pm} - \frac{C}{U_{m, \pm}} + \frac{A + B}{3} \right),
\end{align}
where $m= -1, 0, +1$. 
The spectrum therefore is simply $\{\lambda_0 = \Lambda_{0, +}, \lambda_1 = \Lambda_{1, +}, \, \lambda_2 = \Lambda_{-1, +}, \lambda_3 = A J\}$. The eigenvalue with the largest real part is always $\lambda_0$ whose associated eigenvector \cite{martinez_PRL25} has the following Bloch coordinates 
\begin{equation}
    x^\ts{s} = 0, \quad y^\ts{s} = \frac{2 \lambda_0 J}{4 J^2 + \lambda_0 (\lambda_0 - A J)}, \quad z^\ts{s} = -\frac{\lambda_0 (\lambda_0 - A J)}{4 J^2 + \lambda_0 (\lambda_0 - A J)}.
\end{equation}

\subsection{Complex hopping SDQ}
Consider now the stochastic non-Hermitian Hamiltonian
$$\ha H_t = \frac{J}{\sqrt{2}} (\ha \sigma_x+ \ha \sigma_y) - i (\Gamma + \sqrt{2\gamma} \xi_t) \ket{e}\bra{e}.$$

The antidephasing Liouvillian of this model reads 
\begin{equation}
    \uL^{xy}=J \left(
\begin{array}{cccc}
 0 & e^{+i \frac{3\pi}{4}} & e^{-i \frac{3\pi}{4}} & 0 \\
 e^{+i \frac{\pi}{4}} & A & 0 & e^{-i \frac{3\pi}{4}} \\
e^{-i \frac{\pi}{4}} & 0 & A & e^{+i \frac{3\pi}{4}} \\
 0 & e^{-i \frac{\pi}{4}} & e^{+i \frac{\pi}{4}} & B \\
\end{array}
\right)= J \left(
\begin{array}{cccc}
 0 & (-1)^{\frac{3}{4}} & -(-1)^{\frac{1}{4}} & 0 \\
 (-1)^{\frac{1}{4}} & A & 0 & -(-1)^{\frac{1}{4}} \\
 -(-1)^{\frac{3}{4}} & 0 & A & (-1)^{\frac{3}{4}} \\
 0 & -(-1)^{\frac{3}{4}} & (-1)^{\frac{1}{4}} & B \\
\end{array}\right).
\end{equation}
And its eigenvalues are given by the roots of the characteristic polynomial 
\begin{equation}
    (A-\lambda)(-2 B + (4 +AB)\lambda-(A+B)\lambda^2+ \lambda^3)=0.
\end{equation}
This characteristic polynomial is the same as the one of the real hopping Liouvillian \eqref{eqSM:L}.
This means that all the spectral analysis done for the previous antidephasing Liouvillian in \cite{martinez_PRL25} carries on perfectly to this other system. 
Let us now compute its eigenvectors, particularly that associated with $\lambda_0$, it is given by the solution of the system of equations
\begin{equation}
    \begin{cases}
        e^{+i \frac{3\pi}{4}}b_1+ e^{-i \frac{3\pi}{4}}b_2 = \lambda_0 b_0,\\
        e^{+i \frac{\pi}{4}}b_0+ e^{-i \frac{3\pi}{4}} b_3 = (\lambda_0 - A) b_1,\\
        e^{-i \frac{\pi}{4}}b_0+ e^{+i \frac{3\pi}{4}}b_3 = (\lambda_0 - A) b_2,\\
        e^{-i \frac{\pi}{4}}b_1 + e^{+i \frac{\pi}{4}}b_2 = (\lambda_0 - B) b_3,
    \end{cases}
\end{equation}
which can be solved with Mathematica and gives
\begin{equation}
    |B_0) = \left\{1-\frac{\lambda _0}{B},\frac{e^{i \frac{\pi}{4}} \left(B-2 \lambda _0\right)}{B \left(\lambda _0-A\right)},\frac{e^{i \frac{3 \pi}{4}} \left(B-2 \lambda
   _0\right)}{B \left(A-\lambda _0\right)},\frac{\lambda _0}{B}\right\},
\end{equation}
which has already been normalized such that the trace (sum of first and last element) is unity. 
The Bloch coordinates of the Steady state are
\begin{equation}
    x^\ts{s} = \frac{\sqrt{2}(B - 2 \lambda_0)}{B(\lambda_0-A)}, \quad y^\ts{s} = -\frac{\sqrt{2}(B - 2 \lambda_0)}{B(\lambda_0-A)}, \quad z^\ts{s}= \frac{(B - 2 \lambda_0)}{B}.
\end{equation}

\section{Numerical Solution of Stochastic Differential Equations}

To solve the system of coupled nonlinear SDE's for the Bloch coordinates we use the \textit{explicit order 1.5 strong scheme} by Kloeden and Platen, see Chapt. 11.2 (eqs 2.1, 2.2, 2.3 page 378) of \cite{Kloeden1992}.  For an It\=o SDE of the form
\begin{equation}
    \mr d \be Y = \be a(\be Y) \mr d t + \be b(\be Y) \mr d W_t
\end{equation}
we find its numerical solution by
\begin{align}
&\be Y_{n+1} = \be Y_n + \be b \Delta W + \frac{\be a(\Upupsilon_+){-}\be a(\Upupsilon_-)}{2 \sqrt{\Delta t}}\Delta Z+ \frac{\be a(\Upupsilon_+){+} 2 \be a {+} \be a(\Upupsilon_-)}{4}\Delta t + \frac{\be b(\Upupsilon_+){-}\be b(\Upupsilon_-)}{4\sqrt{\Delta t} }(\Delta W^2 {-} \Delta t) \nonumber \\
&+ \frac{\be b(\Upupsilon_+) {-} 2 \be b {+} \be b(\Upupsilon_-)}{2 \Delta t}(\Delta W \Delta t {-} \Delta Z)+ \frac{\be b(\Phi_+){-}\be b(\Phi_-){-}\be b(\Upupsilon_+)+\be b(\Upupsilon_-)}{4 \Delta t}\left(\frac{\Delta W^2}{3}{-}\Delta t\right)\Delta W,
\end{align}
where $\be a\equiv \be a(\be Y_n), \; \be b \equiv \be b(\be Y_n)$, $\Delta t$ is the time step, the auxiliary variables are
$$\Upupsilon_\pm = \be Y_n + \be a \Delta t \pm \be b \sqrt{\Delta t}, \qquad \Phi_\pm = \Upupsilon_+ \pm \be b(\Upupsilon_+)\sqrt{\Delta t},$$ 
and the noises $\Delta W, \; \Delta Z$ are found from the transformation  (eq. 4.3 pg 352)
$$\Delta W = U_1 \sqrt{\Delta t}, \qquad \Delta Z = \frac{\Delta t^{3/2}}{2}(U_1 + \tfrac{1}{\sqrt{3}}U_2),$$
where $U_1, U_2 \sim \mathcal N(0,1)$ are i.i.d. normal random variables.

\end{document}